\documentclass[journal]{IEEEtran}
\usepackage{color}
\usepackage{epsfig,graphicx}
\usepackage[skip=1ex]{caption}
\usepackage{cite}
\usepackage{subcaption}
\usepackage{amssymb,amsthm}
\usepackage{mathtools, cuted}
\usepackage{lipsum, color, colortbl}
\usepackage[utf8]{inputenc}
\usepackage{cite}
\usepackage{multirow}
\usepackage{array,booktabs}

\usepackage[capitalize]{cleveref}
\usepackage{xr}
\newcolumntype{P}[1]{>{\centering\arraybackslash}p{#1}}
\definecolor{Gray}{gray}{0.9}
\newcommand{\norm}[1]{\left\lVert{#1}\right\rVert}

\newtheorem{theorem}{Theorem}

\newtheorem{lemma}{Lemma}

\title{Information-Theoretic Limits for Steganography in Multimedia}
\author{
	Hassan~Y.~El-Arsh,~
	Amr~Abdelaziz,~\IEEEmembership{Member,~IEEE,}
	Ahmed~Elliethy,~\IEEEmembership{Member,~IEEE,}
	and~Hussein~A.~Aly,~\IEEEmembership{Senior~Member,~IEEE}
	\thanks{H. Y. El-Arsh, Amr Abdelaziz, A. Elliethy, and H. A. Aly are with Dept. of Computer Engineering, Military Technical College, Cairo, Egypt. (e-mail: hassan.yakout@ymail.com, amrashry@mtc.edu.eg, a.s.elliethy@mtc.edu.eg, haly@ieee.org).}
	\thanks{This paper has supplementary downloadable material available at http://ieeexplore.ieee.org, provided by the authors.}
	\thanks{Color versions of one or more of the figures in this paper are available online at http://ieeexplore.ieee.org.}
}

\newcommand{\Sender}{\mathbf{A}}
\newcommand{\Receiver}{\mathbf{B}}
\newcommand{\Eavesdropper}{\mathbf{E}}
\newcommand{\AlphabetCovSteg}{\mathbb{V}}
\newcommand{\AlphabetMsg}{\mathbb{X}}
\newcommand{\AlphabetCodedMsg}{\mathbb{M}}
\newcommand{\CoverObject}{\mathbf{c}}
\newcommand{\StegObject}{\mathbf{s}}
\newcommand{\Msg}{\mathbf{x}}
\newcommand{\CodedMsg}{\mathbf{m}}
\newcommand{\AvgProbErrorEavesdropper}{P_E}
\newcommand{\TotalProbErrorEavesdropper}{P_e}
\newcommand{\ProbErrorReceiver}{P_B}
\newcommand{\CoverDist}{P_c}
\newcommand{\StegDist}{P_s}
\newcommand{\MsgDist}{P_m}
\newcommand{\CoverElement}{c_i}
\newcommand{\StegElement}{s_i}
\newcommand{\CoverElementDist}{P_{c_i}}
\newcommand{\StegElementDist}{P_{s_i}}

\twocolumn

\begin{document}
	
	\maketitle








\begin{abstract}

Steganography in multimedia aims to embed secret data into an innocent multimedia cover object. The embedding introduces some distortion to the cover object and produces a corresponding stego-object. The embedding distortion is measured by a cost function that determines the probability of detection of the existence of secret embedded data. An accurate definition of the cost function and its relation to the maximum embedding rate is the keystone for the proper evaluation of a steganographic system. Additionally, the statistical distribution of multimedia sources follows the Gibbs distribution which is a complex statistical model that prohibits a thorough mathematical analysis.
Previous multimedia steganographic approaches either assume a relaxed statistical distribution of multimedia sources or presume a proposition on the maximum embedding rate then try to prove the correctness of the proposition. Alternatively, this paper introduces an analytical procedure for calculating the maximum embedding rate within multimedia cover objects through a constrained optimization problem that governs the relationship between the maximum embedding rate and the probability of detection by any steganographic detector. In the optimization problem, we use the KL-divergence between the statistical distributions for the cover and the stego-objects to be our cost function as it upper limits the performance of the optimal steganographic detector. To solve the optimization problem, we establish an equivalence between the Gibbs and the correlated-multivariate-quantized-Gaussian distributions for mathematical thorough analysis. The solution to our optimization problem provides an analytical form for the maximum embedding rate in terms of the WrightOmega function. Moreover, we prove that the achieved maximum embedding rate comes in agreement with the well-known square root law (SRL) of steganography. We also establish the relationship between the achieved maximum embedding rate and the experimental results obtained from several embedding and detection steganographic techniques.

\end{abstract}

\section{Introduction}





Steganography aims to embed data within innocent-looking cover objects such as images, audio, video, and even text \cite{steg.types}, to produce a stego-object that looks similar to the original cover object but with hidden data embedded. The embedding process implies some form of distortion to the cover object and this distortion is utilized by a warden to detect if there is hidden data or not using different steganalysis techniques. Consequently, the optimal target for steganography is to hide the maximum amount of data subject to achieving a minimum probability of detection by a warden. In other words, steganography has two competing goals: undetectability and embedding rate.


 


Undetectability is concerned with determining how to alter the cover object to embed data without making a notable distortion to the cover. The distortion is measured by a cost function which is either modeled as the difference between the cover and stego-objects \cite{aly.2011}, or as the expected warden detector sensitivity of changing certain features within the cover object \cite{f5}. On the other hand, the embedding rate is defined as the ratio between the number of hidden data bits to the number of cover data bits with an acceptable undetectability margin \cite[Ch.4]{steg.book}. Maximizing the embedding rate with an acceptable undetectability margin is fundamental to any steganographic algorithm. Multimedia objects are excellent choices for steganography due to their rich structural features and their ubiquity over industry and daily life \cite{jenifer2018survey,tew.2014}.



To evaluate the performance of a multimedia steganographic approach, the relation between the two contradicting goals (undetectability and embedding rate) is a vital measure. Specifically, an accurate definition of the cost function and its relation to the maximum embedding rate is the keystone for proper evaluation of the approach. Additionally, dealing with multimedia incurs additional complexity as the statistical distribution of multimedia sources generally follows the Gibbs distribution \cite{gibbs1984, gibbs1993, bovik3}, which is a complex statistical model. The complexity of the Gibbs distribution prohibits a thorough mathematical analysis, especially when the definition of the cost function requires an accurate definition of the statistical model of the cover object as in \cite[Ch.13]{steg.book} in which the Kullback–Leibler divergence (KL-divergence or the relative entropy), is utilized as a global measure of cover distortion.

Previous approaches of steganography concerned with the definition of the cost function are either not globally accurate for all multimedia types or accurate with limited scope for a certain type of multimedia for a certain class of embedding or detection techniques \cite{quantized.gauss}. Additionally, due to the complexity of Gibbs distribution, previous approaches either use more relaxed mathematical distributions such as Gaussian distribution \cite{limits1,limits2,quantized.gauss,Jessica2015,var.est,Jessica2013,markov2,markov8} in the statistical modeling or presume a preposition on the maximum embedding rate then try to show the correctness of the preposition \cite{limits1, limits2}. Thus the estimated performance may not reflect the real one due to the model relaxation.


The work in\footnote{A preliminary version of this paper with the same authors.} \cite{myconf} partially overcomes the aforementioned problems and proposes a constrained optimization problem that models the undetectability by the KL-divergence and the embedding rate by the mutual information between statistical distributions of the cover and stego-objects. The optimization problem is solved through a rigorous mathematical procedure to present an analytical form for the relation between the embedding rate and undetectability. However, \cite{myconf} utilized a relaxed statistical model for cover and stego-objects (Multivariate-Quantized-Gaussian-Distribution) and does not provide an achievability proof that verifies the reliable extraction of the embedded message at the receiver side when the exact cover realization is not present at the receiver side, in which there exists a probability of decoding error.

In this paper, we present several contributions beyond our preliminary work in~\cite{myconf}. Specifically, we propose a constrained optimization problem that calculates the upper limit that any multimedia steganographic method can achieve reliably (with an achievability proof), for a prespecified acceptable level of detectability by an adversary. Unlike~\cite{myconf}, in this paper, we model the cover object by the Gibbs statistical distribution by establishing an equivalence relation between the Gibbs and the correlated-multivariate-quantized-Gaussian distribution (CMQGD) for rigid mathematical analysis. The solution to our optimization problem provides an analytical form for the maximum embedding rate in terms of the WrightOmega function. Additionally, we prove that the achieved maximum embedding rate comes in agreement with the well-known square root law (SRL) of steganography~\cite[Ch.13]{steg.book}. Finally, we introduce comparisons between our theoretically achieved maximum embedding rate with other practically calculated rates provided by \cite{quantized.gauss}, \cite{markov2}, and \cite{markov8}. The results demonstrate that our calculated upper bound is relatively very small compared to the referenced practical stenographic methods. The reason is that our theoretical upper bound is calculated against the optimal detector, which may not be achieved yet for the referenced steganographic methods.




The paper is organized as follows. Section~\ref{sec:prev} briefly discusses the previous publications related to our work. Section \ref{sec:model} describes the main definitions and assumptions with clear mathematical representation for the proposed constrained optimization problem formulation. Section \ref{sec:MultimediaStoch} provides an equivalence relation between the Gibbs distribution and CMQGD. The solution to the proposed constrained optimization problem is introduced in Section \ref{sec:proof} and the achievability proof is provided in Section \ref{sec:ach}. Section \ref{sec:srl} discusses the relation between the SRL and our results. Practical interpretation of our results is presented in Section \ref{sec:exp} followed by a discussion in Section~\ref{sec:discuss}. The final conclusion is introduced in Section \ref{sec:conc}.

\section{Previous Work}\label{sec:prev}
This section summarizes the most notable prior work related to the scope of this paper. For more details, readers can refer to \cite{steg.book, limits1, limits2, quantized.gauss, var.est, Jessica2013, Jessica2015, markov2, markov8}.

Some of the previous approaches employ the KL-divergence as a statistical cost function and introduce a theoretical limit (under different assumptions) for the maximum number of bits that can be embedded in a designated class of multimedia cover objects under a specified probability of detection by a warden. For instance, \cite{limits1} utilized a special case of Continuous-Gaussian-distributed covers (\textit{AWGN} channels), and introduce additional proofs of achievability and converse. Additionally, \cite{limits2} generalizes the technique used in \cite{limits1} for Multivariate-Continuous-Gaussian-distributed covers (\textit{MIMO} channels). Although these approaches utilized an accurate statistical model for the proposed cover (\textit{AWGN} channels) with a rigorously defined cost function (KL-divergence), their approaches follow a pre-assumed preposition on the maximum embedding rate then try to prove the correctness of the preposition. 

The work in \cite{Jessica2013} provides a systematic technique for constructing the distortion function based on the relation between steganographic Fisher information and KL-divergence. In \cite{Jessica2013}, the additive distortion approach is utilized to model the distortion for digital images, then adaptively calculating the individual pixel costs that minimize the KL-divergence when embedding using the least-significant bit matching approach. Although \cite{Jessica2013} utilizes a simple cover model (zero-mean independent multivariate quantized Gaussian distribution), the achieved security outperforms the state-of-the-art algorithm HUGO \cite{hugo}. The authors improved their work in \cite{Jessica2015} by replacing their previous model in \cite{Jessica2013} with the generalized multivariate Gaussian (referred to as MVGG) combining it with an improved variance estimator. These improvements enhanced the steganographic performance by allowing embedding changes with larger amplitudes in complex (highly textured) regions, thanks to a thicker-tail MVGG model. The MVGG provides comparable performance with respect to pentary coded HiLL \cite{HILL} and S-UNIWARD \cite{SUNIWARD} against maxSRMd2 \cite{maxSRMd2} and SRM \cite{SRM} feature-based steganalysis methods. Although these techniques provide enhanced practical outcomes and an accurate analytical formulation for the relation between the embedding rate and the cost function, their results are based on a relaxed statistical model for the cover.

In \cite{markov2}, the proposed approach follows the non-additive model assumption and utilizes the Gaussian Markov Random Field (GMRF) with four-elements cross neighborhood. The proposed GMRF with low-dimensional clique structures provides the ability to capture the interdependencies among spatially contiguous pixels. The cost function design approach is formulated as the minimization of KL-divergence between the original cover image and the stego one. With GMRF, the cover image is split into two disjoint sub-images, which are conditionally independent. Then, an alternating iterative optimization technique is applied to achieve an efficient embedding incorporated with minimizing the total KL-divergence. The paper provides experiments demonstrating that the performance of GMRF outperforms state-of-the-art steganography MiPOD \cite{var.est} and HiLL techniques in terms of secure payload against SRM and maxSRMd2. This work is expanded in \cite{markov8} with higher-dimension clique structure (eight-elements GMRF). Although \cite{markov2,markov8} provided improved results and an analytical formulation for the relation between the embedding rate and the cost function, the provided results are based on a relaxed statistical model for the cover.

Another approach in \cite{quantized.gauss} focused on gray-scale images as cover objects with an initial assumption that the cover, the message, and the stego-object to be statistically modelled as a Multivariate-Quantized-Gaussian-Distribution. This approach provides a theoretical limit for three popular detection strategies for the likelihood ratio test: Bayes, Minimax, and Neyman-Pearson. This approach maximizes the detection error of these three detectors related to a specified payload. Although this approach provides an accurate analytical form of the relation between the embedding rate and the cost function, the scope of the defined cost function is limited (only three types of detectors) with a relaxed statistical model (Gaussian distribution).

The Square Root Law (SRL)~\cite[Ch.13]{steg.book} is regarded as the first successful work that mathematically models the complex statistical relation between the embedding rate and undetectability in a global framework for multimedia. The SRL states that the embedding rate for any steganographic technique is proportional to the square root of the number of cover elements. For example, assuming cover $A$ contains $x$ elements and cover $B$ contains $4x$ elements. If we can embed $y$ bits of a secret message within $A$ with a probability of detection $\mathcal{P}$ by a certain steganalyzer, then for the same steganalyzer with the same probability of detection $\mathcal{P}$, we can only embed $2y$ bits. Also, for cover $A$, the embedding rate per each cover element will be $\frac{y}{x}$, but for cover $B$ will be $\frac{y}{2x}$. The SRL is valid for any steganographic approach regardless of the embedding and the steganalysis methods. This approach produces only relative approximated order results for any defined cost function without calculating the scaling constant.



\section{System Model and Problem Statement}\label{sec:model}

\subsection{Communication Model} \label{subsec:CM}
\begin{figure}
  \centering
  \resizebox{0.5\textwidth}{!}{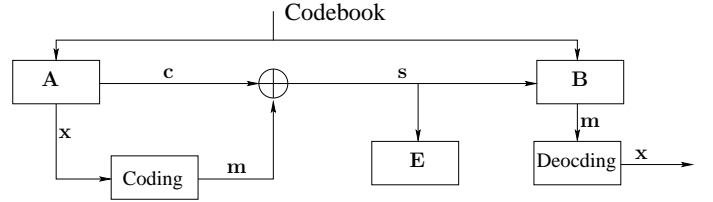}
  \caption{General communication model for steganography.}
  \label{fig:steg}
\end{figure}

\begin{table}
  \centering
  \caption{List of commonly used symbols through this paper.}
  \label{tab.ListOfSymbols}
  \begin{tabular}{ c | l }
  \hline
$\Sender$ & The sender  \\ \hline
$\Receiver$ & The receiver  \\ \hline
$\Eavesdropper$ & The eavesdropper  \\ \hline
$\AlphabetCovSteg$ & Alphabet of cover and stego-elements  \\ \hline
$\AlphabetMsg$ & Alphabet of the original message  \\ \hline
$\AlphabetCodedMsg$ & Alphabet of the coded message  \\ \hline
$\CoverObject$ & The cover object  \\ \hline
$\StegObject$ & The stego-object  \\ \hline
$\Msg$ & The original message  \\ \hline
$\CodedMsg$ & The coded message  \\ \hline
$P_D$ & The probability of steganalyzer correct detection at $\mathbf{E}$  \\ \hline
$\TotalProbErrorEavesdropper$ & The Total probability of steganalyzer error at $\mathbf{E}$  \\ \hline
$\AvgProbErrorEavesdropper$ & The average probability of of steganalyzer error at $\mathbf{E}$ \\ \hline
$\ProbErrorReceiver$ & The probability of decoding error at $\mathbf{B}$ side  \\ \hline
$\CoverDist$ & The joint probability distribution of the cover object \\ \hline
$\StegDist$ & The joint probability distribution of the stego-object  \\ \hline
$\MsgDist$ & The joint probability distribution of the coded message \\ \hline
$\CoverElement$ & The $i^{th}$ cover element  \\ \hline
$\StegElement$ & The $i^{th}$ stego-element  \\ \hline
$\CoverElementDist$ & The probability distribution of the $i^{th}$ cover object  \\ \hline
$\StegElementDist$ & The probability distribution of the $i^{th}$ stego-object  \\ \hline
  \end{tabular}
\end{table}
Elements of a cover object are assumed to be the communication channel between two entities: sender $\Sender$ and its corresponding receiver $\Receiver$. An eavesdropper $\Eavesdropper$ is fully monitoring this communication channel between $\Sender$ and $\Receiver$. Fig.\ref{fig:steg} introduces the general communication model for steganography. Through this paper, we consider an \textit{innocent} original cover $\CoverObject=[c_1, c_2, \dots , c_n]$ and a stego-object $\StegObject=[s_1, s_2, \dots s_n]$ with probability distributions $\CoverDist$ and $\StegDist$, respectively, where $n$ is the number of elements of the cover or the stego-object. Both $\CoverObject,\StegObject \in \AlphabetCovSteg^{n}$, where $\AlphabetCovSteg$ is the set of allowed cover values. For example, when the cover is a gray-scale image, then $\AlphabetCovSteg = \{0,1, \dots ,255\}$ is the set of all available pixel values. It should be noted that $\AlphabetCovSteg$ is not limited to pixels, but can be considered for DC-coefficients \cite{dccoef}, video motion vectors \cite{aly.2011,mvs}, ...etc. Similarly, we consider an original message $\Msg \in \AlphabetMsg^{k}$, where $k$ is the number of elements of the message and $\AlphabetMsg$ is the set of allowed message values (alphabet). The stego-object $\StegObject$ is obtained by embedding a coded message $\CodedMsg \in \AlphabetCodedMsg^{n}$ via an addition process, i.e., $\StegObject = \CoverObject + \CodedMsg$. The coded message $\CodedMsg$ is obtained from $\Msg$ using a certain codebook that maps $\Msg \mapsto \CodedMsg$. The coded message $\CodedMsg$ is statistically distributed as $\MsgDist$.

%


\subsection{Problem Statement}\label{subsec:problem}


Consider the scenario in which $\Sender$ has access to a multimedia cover-object $\CoverObject$, that is drawn from some distribution $\CoverDist$ which is known by both $\Receiver$ and $\Eavesdropper$ . To conceal secret information within $\CoverObject$, $\Sender$ needs to transmit a slightly perturbed version of $\CoverObject$ called the stego-object $\StegObject$ that only $\Receiver$ can reveal the context of these perturbations. Meanwhile, $\Sender$ needs to keep his probability of being detected by $\Eavesdropper$ as minimum as possible.

The fundamental problem here is to quantify, for a given $\CoverDist$, the information-theoretic limits that govern how much information can be exchanged reliably between $\Sender$ and $\Receiver$ while at the same time, maintaining a certain level of probability of their communication being detected by $\Eavesdropper$, namely $P_D$. Since $\Sender$ does not know $\Eavesdropper$'s detection strategy, the transmission strategy should be designed against the optimal detector that $\Eavesdropper$ can deploy. In addition, we assume no prior knowledge of the exact realization of the cover-object neither at $\Receiver$ nor at $\Eavesdropper$. In other words, the cover-object realization is selected at random by $\Sender$ at the time of communication.

%
%

We assume that $\Eavesdropper$ will perform a binary hypothesis test for $\mathcal{H}_0$ and $\mathcal{H}_1$, where $\mathcal{H}_0$ means $\Sender$ is not embedding any secret message whereas $\mathcal{H}_1$ means $\Sender$ does. Thus, we have two types of errors:
\begin{itemize}
  \item Type $\mathbf{I}$ error: Deciding $\mathcal{H}_1$ when $\mathcal{H}_0$ is true, also called false positive. We denote the probability of this type of errors by $\alpha$.
  \item Type $\mathbf{II}$ error: Deciding $\mathcal{H}_0$ when $\mathcal{H}_1$ is true, also called false negative (miss-detection) and we denote the probability of this type of errors by $\beta$.
\end{itemize}

Assuming equal priors for $\Eavesdropper$'s optimal hypothesis test, the total probability of error $\TotalProbErrorEavesdropper$ will be 
\begin{equation}\label{eq.pe_pd}
  \TotalProbErrorEavesdropper = 1- P_D = \alpha + \beta = 2\AvgProbErrorEavesdropper,
\end{equation}
where $\AvgProbErrorEavesdropper$ is the average probability of error for $\Eavesdropper$ under equal priors. According to \cite{lehmann2006testing}, $\TotalProbErrorEavesdropper$ can be calculated as
\begin{equation}\label{eq.pe}
  \TotalProbErrorEavesdropper = 1 - \mathcal{V}(\StegDist,\CoverDist),
\end{equation}
where $\mathcal{V}(\StegDist,\CoverDist)$ is the total variation distance between $\StegDist$ and $\CoverDist$, defined in \cite{lehmann2006testing} as
\begin{equation}\label{eq.v}
  \mathcal{V}(\StegDist,\CoverDist) = \dfrac{1}{2}\norm{\StegDist-\CoverDist}_1,
\end{equation}
where $\norm{.}_1$ is the $L_1$ norm. As $L_1$ norm's analytic calculations are not wieldy, $\mathcal{V}(\StegDist,\CoverDist)$ is related to the KL-divergence as \cite[Ch.11]{info.th.book}:
\begin{equation}\label{eq.klv}
  \mathcal{V}(\StegDist,\CoverDist) \leq \sqrt{\dfrac{1}{2} \mathcal{D}(\StegDist \parallel \CoverDist)},
\end{equation} 
where the KL-divergence is given by:
\begin{align}\label{eq.kl}
 \mathcal{D}(\StegDist \parallel \CoverDist) & = \sum_{j=1}^n P_{s_j} \ln \frac{P_{s_j}}{P_{c_j}}.
\end{align} 
where $P_{s_j}$ and $P_{c_j}$ are the bins of probability mass function for cover and stego-objects, respectively.
For $\Sender$ to guarantee a low detection probability at $\Eavesdropper$'s optimal detector, $\Sender$ needs to bound $\mathcal{V}(\StegDist,\CoverDist)$ by some $\epsilon$ chosen according to the desired probability of $\Eavesdropper$'s detection. Consequently, $\Sender$ ensures that the sum of error probabilities at $\Eavesdropper$ is bounded as $\alpha + \beta = 1-\mathcal{V}(\StegDist,\CoverDist) $. Using (\ref{eq.klv}), $\Sender$ can achieve this goal by designing a steganographic technique such that
\begin{equation}
\label{eq.c1}
\mathcal{D}(\StegDist \parallel \CoverDist) \leq 2\epsilon^2.
\end{equation} 

On the other hand, $\Receiver$ must be able to correctly decode the message $\Msg$ from $\CodedMsg$ sent by $\Sender$. Thus the chosen codebook must be selected such that $\MsgDist$ maximizes the mutual information between both $\StegDist$ and $\MsgDist$, which is defined as 
\begin{equation}\label{eq.mi}
  I(\MsgDist ; \StegDist) = H(\MsgDist) - H(\MsgDist\mid \StegDist),
\end{equation}
where $H(\MsgDist)$ is the entropy of $\MsgDist$ and $H(\MsgDist\mid \StegDist)$ is the conditional entropy of $\MsgDist$ given $\StegDist$. Thus, our constrained optimization problem can be written as
\begin{equation}\label{eq.constrain_org}
  \underset{\MsgDist}{\mathrm{argmax}} \quad I(\MsgDist ; \StegDist) \qquad  \mathrm{s.t.} \qquad  \mathcal{D}(\StegDist \parallel \CoverDist) \leq 2\epsilon^2.
\end{equation}
Note that, $I(\MsgDist ; \StegDist)$ in (\ref{eq.mi}) can be written as~\cite[Ch.2]{info.th.book}
\begin{equation}\label{eq.mi2}
  I(\MsgDist ; \StegDist) = H(\StegDist) - H(\StegDist\mid \MsgDist) = H(\StegDist) - H(\CoverDist), 
\end{equation}
and because $H(\CoverDist)$ is given, maximizing $I(\MsgDist ; \StegDist)$ is equivalent to maximizing $H(\StegDist)$. Thus, the optimization problem in \eqref{eq.constrain_org} can be restated as 
\begin{equation}\label{eq.constrain}
 \boxed{ \underset{\MsgDist}{\mathrm{argmax}} \quad H(\StegDist) \qquad  \mathrm{s.t.} \qquad  \mathcal{D}(\StegDist \parallel \CoverDist) \leq  2\epsilon^2}.
\end{equation}

\section{Multimedia stochastic model}\label{sec:MultimediaStoch}

In this section, a mathematical illustration for our basic assumptions about the general statistical distribution model for the multimedia ($\CoverDist$) is introduced. Consequently, for a clear illustration of the multimedia stochastic model, we need first to define the Markov-Random-Field (MRF) and Gibbs distribution. 

MRF is a multidimensional random process, which generalizes the single dimensional Markov random process \cite{bovik3}. Let $\ddot{X}$ be a coordination system in $R^N$ and $\rho(i)$ is a function representing the neighbourhood for each element $i\in \ddot{X}$, such that $i \notin \rho(i)$ and $i \in \rho(j)$ \textit{iff} $j \in \rho(i)$. For example, the neighbourhood may be defined as the immediate left, right, top and bottom neighbours of $i$. Let  $\ddot{Y}$ be the neighbourhood system representing the set of neighbourhoods of all elements $i,j,\hdots \in \ddot{X}$.

A random field $\xi$ over $\ddot{X}$ is a multidimensional random process where each element $i\in \ddot{X}$ is assigned a random variable $\xi_i$ with $\mathfrak{f}_i$ be its associated realization for all $i \in \ddot{X}$. $\xi$ is called a MRF if it achieves the following conditions:
\begin{itemize}
\item Positivity property: $P(\xi=\mathfrak{f}) > 0$, $\forall \mathfrak{f} \in \ddot{S}$, and
\item Markovianity property: $P(\xi_i=\mathfrak{f}_i | \xi_j=\mathfrak{f}_j , \forall j \neq i) = P(\xi_i=\mathfrak{f}_i | \xi_j=\mathfrak{f}_j , \forall j \in \rho(i) ), \forall i \in \ddot{X}, \forall \mathfrak{f} \in \ddot{S}$,
\end{itemize}
where $P$ is the probability measure and $\ddot{S}$ is the state space for the MRF $\xi$.

According to \cite{bovik3}, Hammersley-Clifford theorem states that $\xi$ is a MRF over $\ddot{X}$ with respect to $\ddot{Y}$ if and only if its probability distribution follows the Gibbs distribution with respect to $\ddot{X}$ and $\ddot{Y}$. To explain the Gibbs distribution, the idea of \textit{clique} must to be clarified. A clique $\omega$ is a correlated group of neighboured elements, where $\omega\subset\ddot{X}$ with respect to $\ddot{Y}$ (i.e.: $\omega$ consists of a single element $i$ or multiple elements $i,j,\hdots$ which are neighbours). Also, $\omega \in \ddot{\Omega}$, where $\ddot{\Omega}$ denotes the set of all cliques.

According to \cite{gibbs1984, gibbs1993, bovik3}, it is accurate to statistically model the multimedia objects by the Gibbs distribution, which implies the above described properties (positivity and markovianity), in which all elements for each multimedia object are organized in \textit{cliques}. Gibbs distribution quantifies the probability $P(\xi=\mathfrak{f})$ through an energy function $U_\mathfrak{f}$ and a temperature constant $\hat{T}$ as
\begin{equation}
P(\xi=\mathfrak{f}) = \frac{1}{Z} e^{- U_\mathfrak{f} / \hat{T}},
\label{eq.gibbs}
\end{equation}
where 
\begin{equation}
Z = \sum_{k=1}^{M}{e^{- U_k / \hat{T}}},
\end{equation}
is the normalization denominator that is called the \textit{partition function} and $M$ is the number of all allowed states of the system (number of elements within $\ddot{S}$). The energy function is modelled as
\begin{equation}\label{eq.gibbs.energy.func}
U_\mathfrak{f} = \sum_{\omega\in \ddot{\Omega}} V_\mathfrak{f}(\omega),
\end{equation}
where $V_\mathfrak{f}(\omega)$ is the potential function for $\mathfrak{f}$ calculated only within the clique $\omega \in \ddot{\Omega}$.

Gibbs distribution involves computing the partitioning function which is intractable for most models. Therefore, in this paper, we propose an efficient approximation for the Gibbs distribution by modelling it with the \textit{correlated-multivariate-quantized-Gaussian-distribution} (CMQGD) for more thorough mathematical analysis. One possible approximation to \eqref{eq.gibbs} is to model $V_f(\omega)$ in \eqref{eq.gibbs.energy.func} as
\begin{equation}\label{eq.PotFun}
V_\mathfrak{f}(\omega) = \frac{\hat{T}}{2}(\mathfrak{f}_\omega - \mu_\omega)\Sigma_\omega^{-1}(\mathfrak{f}_\omega - \mu_\omega)^T, 
\end{equation}
where
\begin{itemize}
\item $\mathfrak{f}_\omega$ is the realization of the random field $\xi$ within the clique $\omega$ (i.e.: $\mathfrak{f}_i,\mathfrak{f}_j \in \mathfrak{f}_\omega \quad \forall i,j,\hdots \in \omega$ and $\mathfrak{f}_\omega \subset \mathfrak{f}$ ).
\item $\mu_\omega$ and $\Sigma_\omega$ are the mean vector for $\xi_\omega \subset \xi$ and  its covariance matrix within the clique $\omega$, respectively.
\end{itemize}
This approximation is valid for the following reasons:
\begin{itemize}
\item $V_\mathfrak{f}(\omega)$ can be mapped to any function depends only on the elements within the clique $\omega$ \cite{bovik3}.
\item The multiplication of $(\mathfrak{f}_\omega - \mu_\omega)$ and $(\mathfrak{f}_\omega - \mu_\omega)^T$, which is the distance from the mean calculated locally within the clique $\omega$, controls the probability of certain realization (state) $\mathfrak{f}_\omega$ (and hence $P(\xi=\mathfrak{f})$) within the clique $\omega$ for certain $\Sigma_\omega$. In other words, the local coherency condition of multimedia elements: $P(\xi=\mathfrak{f}) \propto -U_\mathfrak{f} \propto -V_\mathfrak{f}(\omega)$ is achieved. This allows lower energy states to be always have a higher probability than the higher energy ones, which is one of the main Gibbs properties \cite{gibbs1984}.
\item $Z$, which is the normalization denominator, is equivalent to $\sqrt{2\pi}|\Sigma|$ as $\sqrt{2\pi}|\Sigma| = \sum_{\omega\in \ddot{\Omega}} e^{-\sum{\frac{\hat{T}}{2}(\mathfrak{f}_\omega - \mu_\omega)\Sigma_\omega^{-1}(\mathfrak{f}_\omega - \mu_\omega)^T}}$. 
\end{itemize}
It should be noted that other valid assumptions for the potential function $V_\omega(\mathfrak{f})$ can be used to approximate the Gibbs distribution with other distributions. The approximation in \eqref{eq.PotFun} is one of them.

%

\section{Main Results}\label{sec:proof}
In this section, we discuss the solution and results of the optimization problem~\eqref{eq.constrain} in Sec.~\ref{sec:model} in subsection \ref{subsec.solution} using the assumptions in subsection \ref{subsec.assumptions}.
\subsection{The main assumptions}\label{subsec.assumptions}
To solve the optimization problem~\eqref{eq.constrain}, we have the following assumptions:
\begin{itemize}
\item From here on, each clique within the cover will be considered as a single cover element $\CoverElement$ with \textit{unknown distribution} $\CoverElementDist$.
\item Having established the equivalence relation between Gibbs and Gaussian distributions as discussed in section \ref{sec:MultimediaStoch}, we model the whole cover object (\textit{i.e.}: the joint distribution for the cover object $\CoverDist$) as a CMQGD with mean $\vec{\mu_c}$ and covariance matrix $\Sigma_c$. Rather, we have no assumptions on the marginal distribution $\CoverElementDist$ of each cover element $\CoverElement$.
\item The cover statistical parameters ($\vec{\mu_c}$, $\Sigma_c$) are known to all $\Sender$, $\Receiver$, and $\Eavesdropper$.
\item Although we have no assumptions for both $\StegDist$ and $\MsgDist$, according to \cite{gaussisgauss}, to minimize the KL-divergence for any Gaussian distributed cover object, the corresponding stego-object distribution must be Gaussian. Thus, the embedding operation will produce another CMQGD $\StegDist$ with parameters ($\vec{\mu_s}$, $\Sigma_s$).
\item The realization of codebook from $\MsgDist$ (the exact codewords used between $\Sender$ and $\Receiver$) are shared only between $\Sender$ and $\Receiver$ and not known to $\Eavesdropper$.
\end{itemize}

\subsection{The Solution to the optimization problem in \eqref{eq.constrain}}\label{subsec.solution}
To solve~\eqref{eq.constrain}, we need to determine $H(\CoverDist)$, $H(\StegDist)$ and the KL-Divergence $\mathcal{D}(\StegDist \parallel \CoverDist)$. Using lemma \ref{lem.A1} in appendix (A), $H(\CoverDist)$ and $H(\StegDist)$ are computed as demonstrated in \eqref{eq.ent.pc} and \eqref{eq.ent.ps}, respectively. The KL-Divergence between two quantized distributions can be evaluated using \eqref{eq.kl.epsilon} alongside with the following lemma. 
\begin{lemma}\label{TH1}
According to \cite[Ch.11]{info.th.book}, KL-Divergence between any two uniformly quantized distributions $F(\dot{X})$ and $G(\dot{X})$ is bounded as:
	\begin{equation}\label{eq.th1}
	\mathcal{D}(F(\dot{X}) \parallel G(\dot{X})) \leq \mathcal{D}(f(\dot{x}) \parallel g(\dot{x}))
	\end{equation}
	Where $f$ and $g$ are the continuous versions of $F$ and $G$, respectively, and $\dot{X}$ is the uniformly quantized version of the continuous random variable $\dot{x}$.
\end{lemma}

From lemma~\ref{TH1} and the definition of the KL-divergence for multivariate-continuous-Gaussian-distributions~in \cite{limits2}, the KL-Divergence for CMQGD can be bounded as
\begin{align}\label{eq.kl:qgd}
\mathcal{D}(\StegDist \parallel \CoverDist) & \leq \frac{1}{2} \bigg(\mathrm{tr}(\Sigma_c^{-1} \Sigma_s) + (\vec{\mu_c} - \vec{\mu_s})^T \Sigma_c^{-1} (\vec{\mu_c} - \vec{\mu_s}) \nonumber \\
& + \ln \frac{|\Sigma_c|}{|\Sigma_s|} - n\bigg),
\end{align} 
where $n$ is number of cover elements. For more conservative evaluation for our optimization problem in \eqref{eq.constrain}, we set $\mathcal{D}(\StegDist \parallel \CoverDist)$ to its upper bound in the mathematical relations in the rest of the paper. Specifically,
\begin{align}\label{eq.kl.epsilon}
\Aboxed{\mathcal{D}(\StegDist \parallel \CoverDist) & = \frac{1}{2} \bigg(\mathrm{tr}(\Sigma_c^{-1} \Sigma_s) + (\vec{\mu_c} - \vec{\mu_s})^T \Sigma_c^{-1} (\vec{\mu_c} - \vec{\mu_s})} \\ \nonumber
\Aboxed{& + \ln \frac{|\Sigma_c|}{|\Sigma_s|} - n\bigg)}.
\end{align}

Having established the upper bound of of the KL-Divergence for CMQGD in \eqref{eq.kl.epsilon}, the solution of the optimization problem in~\eqref{eq.constrain} as provided in the following theorem. 
\begin{theorem}\label{TH2}
With the assumptions stated in subsection \ref{subsec.assumptions}, the solution of \eqref{eq.constrain} can be achieved by setting: 
\begin{enumerate}
\item The distribution of $\MsgDist$ must be CMQGD (same as cover and stego-objects), with the parameters (mean $\vec{\mu_m}$ and variance $\Sigma_m$) as
	\begin{enumerate}
	\item $\vec{\mu_m} = 0$.
	\item $\Sigma_m = \Big( -W(-\frac{4\epsilon^2}{n} -1 -i\pi) -1 \Big)\Sigma_c$.
	\end{enumerate}
	for optimal steganalysis detector with detection capabilities limited by $\epsilon$ (i.e.: $P_D \leq \sqrt{\frac{\mathcal{D}(\StegDist \parallel \CoverDist)}{2}} \leq \epsilon)$, where $W(.)$ is the WrightOmega function \cite{WrightOmega}.
\item The maximum achievable embedding rate $\boxed{I(\StegDist;\MsgDist)$ is $\frac{n}{2}\ln(-W(-\frac{4\epsilon^2}{n} -1 -i\pi))}$ for sufficiently large $n$.
\end{enumerate} 
 \end{theorem}
\textbf{\textit{Proof}}: 

The Lagrangian $\mathcal{L}$ of \eqref{eq.constrain} will be 
\begin{equation}\label{eq.lag}
\mathcal{L}(\StegDist, \lambda) = \mathcal{L}(\vec{\mu_s}, \Sigma_s, \lambda) = H(\StegDist) - \lambda(\mathcal{D}(\StegDist \parallel \CoverDist) - 2\epsilon^2),
\end{equation}
where $\lambda$ is the Lagrange multiplier. To find the optimized parameters for \eqref{eq.lag}, we set $\frac{\partial}{\partial \vec{\mu_s}}\mathcal{L} = 0$ and $\frac{\partial}{\partial \Sigma_s}\mathcal{L} = 0$. With the definitions of $H(\StegDist)$ and $\mathcal{D}(\StegDist \parallel \CoverDist)$ in \eqref{eq.ent.ps} and \eqref{eq.kl.epsilon}, respectively, we have
\begin{align}
\label{eq.start1} \frac{\partial}{\partial \vec{\mu_s}} H(\StegDist) & =  0 , \\
\label{eq.start2} \frac{\partial}{\partial \Sigma_s} H(\StegDist) & \approx  \frac{1}{2}(\Sigma_s^{-1})^T ,\\
\label{eq.start3.5} \frac{\partial}{\partial \vec{\mu_s}} \mathcal{D}(\StegDist \parallel \CoverDist) & =  \frac{-1}{2}\Big(\Sigma_c^{-1} (\vec{\mu_c} - \vec{\mu_s}) + (\Sigma_c^{-1})^T (\vec{\mu_c} - \vec{\mu_s}) \Big), \\
\label{eq.start4} \frac{\partial}{\partial \Sigma_s} \mathcal{D}(\StegDist \parallel \CoverDist) & = \frac{1}{2} \big( \Sigma_c^{-1} - \Sigma_s^{-1} \big).
\end{align}
As $\Sigma_c$ and $\Sigma_s$ are symmetric matrices, \eqref{eq.start3.5} can be rewritten as
\begin{align}
\label{eq.start3} \frac{\partial}{\partial \vec{\mu_s}} \mathcal{D}(\StegDist \parallel \CoverDist) & =  -\Sigma_c^{-1} (\vec{\mu_c} - \vec{\mu_s}).
\end{align}
From \eqref{eq.start1} and \eqref{eq.start3}, as $\frac{\partial}{\partial \vec{\mu_s}}\mathcal{L} = 0$, we have
\begin{equation}\label{eq.result1}
\vec{\mu_s} = \vec{\mu_c}.
\end{equation}
From \eqref{eq.start2} and \eqref{eq.start4}, as $\frac{\partial}{\partial \Sigma_s}\mathcal{L} = 0$, we have
\begin{align}\label{eq.result2.a}
(\lambda +1) \Sigma_s^{-1} &\approx \lambda \Sigma_c^{-1}, \nonumber \\
\Sigma_s &\approx \frac{\lambda +1}{\lambda} \Sigma_c, \nonumber \\ 
\Sigma_s &\approx a \Sigma_c.
\end{align}
Through the paper, we name the factor $a = \frac{\lambda +1}{\lambda}$ as the \textit{embedding factor}. From~\eqref{eq.result2.a}, it is clear that $\Sigma_s$ has the exact eigenvectors as $\Sigma_c$ and the eigenvalues of $\Sigma_s$ are a scaled version of the eigenvalues of $\Sigma_c$ with the embedding factor $a$. 
Let $\psi_{c}(\vec{t})$ and $\psi_{s}(\vec{t})$ be the characteristic functions for both the cover and the stego-objects where
\begin{align*}
\psi_{c}(\vec{t}) = e^{i\vec{\mu_c}^T \vec{t} + \frac{1}{2}\vec{t}^T\Sigma_c \vec{t}}, \\
\psi_{s}(\vec{t}) = e^{i\vec{\mu_s}^T \vec{t} + \frac{1}{2}\vec{t}^T\Sigma_s \vec{t}}.
\end{align*}
As $\StegObject = \CoverObject + \CodedMsg$, this implies $\StegDist = \CoverDist \oplus \MsgDist$, where $\oplus$ represents the convolution operation. Thus, the characteristic function for the distribution of codebook of the message ($\psi_{m}(\vec{t})$) will be
\begin{align*}
\psi_{m}(\vec{t}) = & \frac{\psi_{s}(\vec{t})}{\psi_{c}(\vec{t})} = e^{i(\vec{\mu_s}^T-\vec{\mu_c}^T)\vec{t}+ \frac{1}{2}\vec{t}^T (\Sigma_s - \Sigma_c) \vec{t}}.
\end{align*}
Then from \eqref{eq.result1}, we have
\begin{equation}\label{eq.ch_m}
	\psi_{m}(\vec{t}) = e^{ \frac{1}{2}\vec{t}^T(\Sigma_s - \Sigma_c) \vec{t}} = e^{ \frac{1}{2}\vec{t}^T(\Sigma_m) \vec{t}}.
\end{equation}
Thus
\begin{equation} \label{eq.mum2}
\boxed{\mu_m = 0}.
\end{equation}
This means that, the codebook of the message must be modelled as CMQGD with zero mean and variance
\begin{align}
\Sigma_m \approx & \Sigma_s - \Sigma_c  \nonumber \\
				= & \boxed{( a -1 )\Sigma_c}.\label{eq.sigmam2}
\end{align}

The embedding factor $a$ can be calculated as follows. Starting from \eqref{eq.result2.a}, we have
\begin{equation}\label{eq.result2.b}
\mathrm{tr}(\Sigma_c^{-1} \Sigma_s) \approx na,
\end{equation}
and 
\begin{equation}\label{eq.result2}
\ln \frac{|\Sigma_c|}{|\Sigma_s|} \approx -n \ln (a).
\end{equation}
Substituting from \eqref{eq.result1}, \eqref{eq.result2.b} and \eqref{eq.result2} into \eqref{eq.kl.epsilon}, we get
\begin{align}\label{eq.result3}
a n - n \ln(a) &\approx 2 \mathcal{D}(\StegDist \parallel \CoverDist) +n, \nonumber \\
a -\ln(a) &=\approx\frac{2}{n} \mathcal{D}(\StegDist \parallel \CoverDist) +1.
\end{align}
The solution of \eqref{eq.result3} can be obtained from \cite{WrightOmega} as
\begin{equation}\label{eq.result4}
a \approx -W(-\frac{2}{n} \mathcal{D}(\StegDist \parallel \CoverDist) -1 -i\pi).
\end{equation}
From \eqref{eq.result4} and according to \cite{WrightOmega}, $a$ is monotonically increasing with $\mathcal{D}(\StegDist \parallel \CoverDist)$. From \eqref{eq.result2.a} and \eqref{eq.ent.ps}, $a$ is monotonically increasing with $H(\StegDist)$. Thus, using the constraint in \eqref{eq.constrain} by substituting $\mathcal{D}(\StegDist \parallel \CoverDist)$ with $2\epsilon^2$ in \eqref{eq.lag} we have
\begin{equation}\label{eq.result4m}
\boxed{a^* =-W(-\frac{4\epsilon^2}{n} -1 -i\pi)},
\end{equation}
where $a^* \in \mathbb{R}$ is the embedding factor\footnote{We proved in Lemma~\ref{lem.A*} that $a^* \in \mathbb{R}$.} associated with the design parameter $\epsilon$. Thus, \eqref{eq.mum2}, \eqref{eq.sigmam2} and \eqref{eq.result4m} complete the proof of the first part of  Theorem~\ref{TH2}. $\blacksquare$

From \eqref{eq.mi2}, we have:
\begin{align}
I(\StegDist;\MsgDist)	&=		  H(\StegDist) - H(\CoverDist).		\nonumber
\end{align}
Then from \eqref{eq.ent.pc} and \eqref{eq.ent.ps}:
\begin{align}
I(\StegDist;\MsgDist) &\approx  \frac{1}{2}\ln(2\pi e|\Sigma_s|)- \frac{1}{2}\ln(2\pi e|\Sigma_c|). \nonumber
\end{align}
Then, from \eqref{eq.sigmam2} and \eqref{eq.result4m}:
\begin{align}\label{eq.I} 
I(\StegDist;\MsgDist)	&\approx \frac{n}{2}\ln(a^*) + \frac{1}{2}\ln(2\pi e|\Sigma_c|)- \frac{1}{2}\ln(2\pi e|\Sigma_c|) \nonumber \\
			&=		  \frac{n}{2}\ln(a^*). 
\end{align}
Thus
\begin{equation}\label{eq.main}
I(\StegDist;\MsgDist) \approx \frac{n}{2}\ln\Big(-W(-\frac{4\,\epsilon^2}{n} -1 -i\pi)\Big).
\end{equation}
Equation \eqref{eq.main}\footnote{It should be noted that for continuously distributed covers such as \cite{limits1} and \cite{limits2}, the approximations in \eqref{eq.th1}, \eqref{eq.ent.pc}, and \eqref{eq.ent.ps} will be changed to equality. Consequently, for such cases, \eqref{eq.main} will be valid with equality.} completes the proof of the second part of  Theorem~\ref{TH2}. $\blacksquare$

Please note that, despite $\mathcal{D}(\StegDist \parallel \CoverDist) \neq \mathcal{D}(\CoverDist \parallel \StegDist)$, the results obtained from Theorem \ref{TH2} are also valid for $\mathcal{D}(\CoverDist \parallel \StegDist)$. The proof of this claim in provided in the following lemma.

\begin{lemma}\label{TH3}
$\vec{\mu_m}$ and $\Sigma_m$ obtained from Theorem \ref{TH2} are also valid for $\mathcal{D}(\CoverDist \parallel \StegDist)$.
\end{lemma}
\textbf{\textit{Proof:}} Equation \eqref{eq.kl.epsilon} can be modified as:
\begin{align}\label{eq.kl:qgd.kl2}
\mathcal{D}(\CoverDist \parallel \StegDist) & = \frac{1}{2} \bigg(\mathrm{tr}(\Sigma_s^{-1} \Sigma_c) + (\vec{\mu_s} - \vec{\mu_c})^T \Sigma_s^{-1} (\vec{\mu_s} - \vec{\mu_c}) \nonumber \\
& + \ln \frac{|\Sigma_s|}{|\Sigma_c|} - n\bigg).
\end{align}
Consequently, \eqref{eq.start4} and \eqref{eq.start3} will be modified as:
\begin{equation}\label{eq.start3.kl2}
\frac{\partial}{\partial \Sigma_s} \mathcal{D}(\CoverDist \parallel \StegDist)  = \frac{1}{2} \big( -(\Sigma_s^{-1})^T \Sigma_c (\Sigma_s^{-1})^T + (\Sigma_s^{-1})^T \big),
\end{equation}
\begin{equation}\label{eq.start4.kl2}
\frac{\partial}{\partial \vec{\mu_s}} \mathcal{D}(\CoverDist \parallel \StegDist) =  \Sigma_s^{-1} (\vec{\mu_s} - \vec{\mu_c}).
\end{equation}

Following the same steps of the proof of Theorem \ref{TH2} using \eqref{eq.start1}, \eqref{eq.start2}, \eqref{eq.start3.kl2}, and \eqref{eq.start4.kl2}, the results presented at \eqref{eq.mum2}, \eqref{eq.sigmam2}, \eqref{eq.result4}, and \eqref{eq.result4m} will be the same as Theorem \ref{TH2}. $\blacksquare$


%

\section{Achievability}\label{sec:ach}
In this section, we prove that the maximum embedding rate $I(\StegDist;\MsgDist)$ in~\eqref{eq.main} is achievable with a low probability of decoding error $\ProbErrorReceiver$ at $\Receiver$. We couldn't include the achievablity constraint in the optimization problem in \eqref{eq.constrain} and \eqref{eq.lag} as $\StegDist$ and $\MsgDist$ will not be obtained till solving the optimization problem.

We utilize the \textit{standard random coding argument} \cite[Ch.7]{info.th.book} as a base for our proof. Although the minimum distance decoder (maximum likelihood estimator) is the optimal detector~\cite[Ch.7]{info.th.book}, there is technical difficulties in using this detector to calculate $\ProbErrorReceiver$ as we do not have the marginal distribution for each cover element; only the joint distribution for the whole cover. Thus, we utilize the \textit{jointly typical decoder} \cite[Ch.7]{info.th.book} at $\Receiver$ instead.

In our achievability proof, we utilize the same procedures for the Channel Capacity Theorem (Theorem (9.1.1) in \cite{info.th.book}) for calculating $\ProbErrorReceiver$. Although this theorem mandates the constraint for the average transmission power, in our case we do not need this power constraint as $\Sigma_m$ is very small due to the constraint of low probability of detection at $\Eavesdropper$ in \eqref{eq.constrain}. Without loss of generality, assuming transmitting the $i^{th}$ codeword, we have two types of errors
\begin{itemize}
\item The transmitted codeword and the received one are not jointly typical. We denote this error by $\hat{E}_i$.
\item The received codeword is jointly typical with another wrong non-intended codeword. This error is denoted by $\sum_{j=1, j\ne i}^{\mathcal{K}} E_j$, where $\mathcal{K}$ is the number of all codewords.
\end{itemize}
Thus, equation (9.26) in \cite{info.th.book} will be modified as
\begin{align}
\ProbErrorReceiver \leq P(\hat{E}_i) + \sum_{j=1, j\ne i}^{\mathcal{K}} P(E_j),
\end{align}
where $P(x)$ is the probability of an event $x$. Continuing the same procedures for the proof of Theorem (9.1.1) in \cite{info.th.book}: 
\begin{equation}\label{eq.pb}
\ProbErrorReceiver \leq 2\delta,
\end{equation}
where $\delta >0$ \cite[Ch.3]{info.th.book} is an arbitrary small number. Thus, \eqref{eq.pb} prove that when $n$ is sufficiently large and for an actual embedding rate $R\leq I(\StegDist;\MsgDist)-2\epsilon$, then $\ProbErrorReceiver$ is sufficiently small. This proves the existence of a codebook ($\mathcal{K}$,$n$,$\epsilon$) that achieves an embedding rate $R\leq I(\StegDist;\MsgDist)-2\epsilon$ at $\Receiver$ with low $\ProbErrorReceiver$.

It should be noted that although related approaches such as \cite{limits1} and \cite{limits2} utilize a converse proof, in our case we have dropped the converse proof as we have got the maximum embedding rate as a solution for the optimization problem in \eqref{eq.constrain}, which guarantees that there is no other codebook generated by any other distribution can achieve higher rate than \eqref{eq.main}.



\section{Relation to the square root law of steganography}\label{sec:srl}
In this section, we establish the relation between the obtained expressions in \eqref{eq.result4m} and \eqref{eq.main} in terms of the WrightOmega function and the well known previously obtained results of the SRL in \cite{steg.book,limits1,limits2}. To do so, we state the following facts about the WrightOmega function:
\begin{itemize}
\item As the WrightOmega function in the form $-W(-\gamma-1-i\pi)$ is monotonically increasing with $\gamma$; where $\gamma$ is a positive constant; $-W(-\gamma-1-i\pi)$ can be approximated by polynomial regression in the form
\begin{equation}\label{eq.Wtheta}
a^* = -W(-\gamma-1-i\pi) = 1+\theta_1 \gamma + \theta_2 \gamma^2 + \theta_3 \gamma^3 + \hdots, 
\end{equation}
 where $\theta_i$ is an arbitrary constant. 
\item As $\gamma = \frac{4\epsilon^2}{n}$, then for large $n$ we can ignore the higher order terms in \eqref{eq.Wtheta} and use the following approximation\footnote{It should be noted that $\theta_1$ must be a positive constant according to \eqref{eq.agt1}.}
\begin{equation}\label{eq.a.approx.1}
a^* \approx 1 + \theta_1\frac{4\epsilon^2}{n}.
\end{equation} 
\item From Lemma \ref{lemma.a.geq.1}, as $a^*\geq 1$, then $\sqrt{a^*} \geq 1$. Also, as $(1+x)^{\frac{1}{2}}\leq 1+x^\frac{1}{2}$ and $\ln(1+x) \leq x$, then the following approximation of \eqref{eq.I} is valid
\begin{align}\label{eq.Isrl}
I(\StegDist;\MsgDist) &\approx n\ln (\sqrt{a^*}) \nonumber \\
  &\approx n\ln\Big(\sqrt{1+\frac{4\epsilon^2}{n}}\Big)  \nonumber \\
  &\leq n\ln\big(1+\frac{2\epsilon}{\sqrt{n}}\big) \nonumber \\
  &\leq n\frac{2\epsilon}{\sqrt{n}}, \nonumber \\
\Aboxed{I(\StegDist;\MsgDist) &\leq 2\epsilon\sqrt{n}}.
\end{align}
\end{itemize}
Thus, \eqref{eq.Isrl} establishes the relation between \eqref{eq.main} and the SRL. Hence, \eqref{eq.Isrl} proves that our results come in agreement with previously established SRL in \cite{steg.book} in the context of image steganography and \cite{limits1,limits2} in the context of covert communication. Using the justifications in section \ref{sec:MultimediaStoch} for approximating the Gibbs distribution with CMQGD  and the proved relation to the SRL in \eqref{eq.Isrl}, equation \eqref{eq.main} can be considered as a rigorous analytic form for the SRL.~$\blacksquare$ 

\begin{figure}
  \centerline{\includegraphics[width=0.75\linewidth]{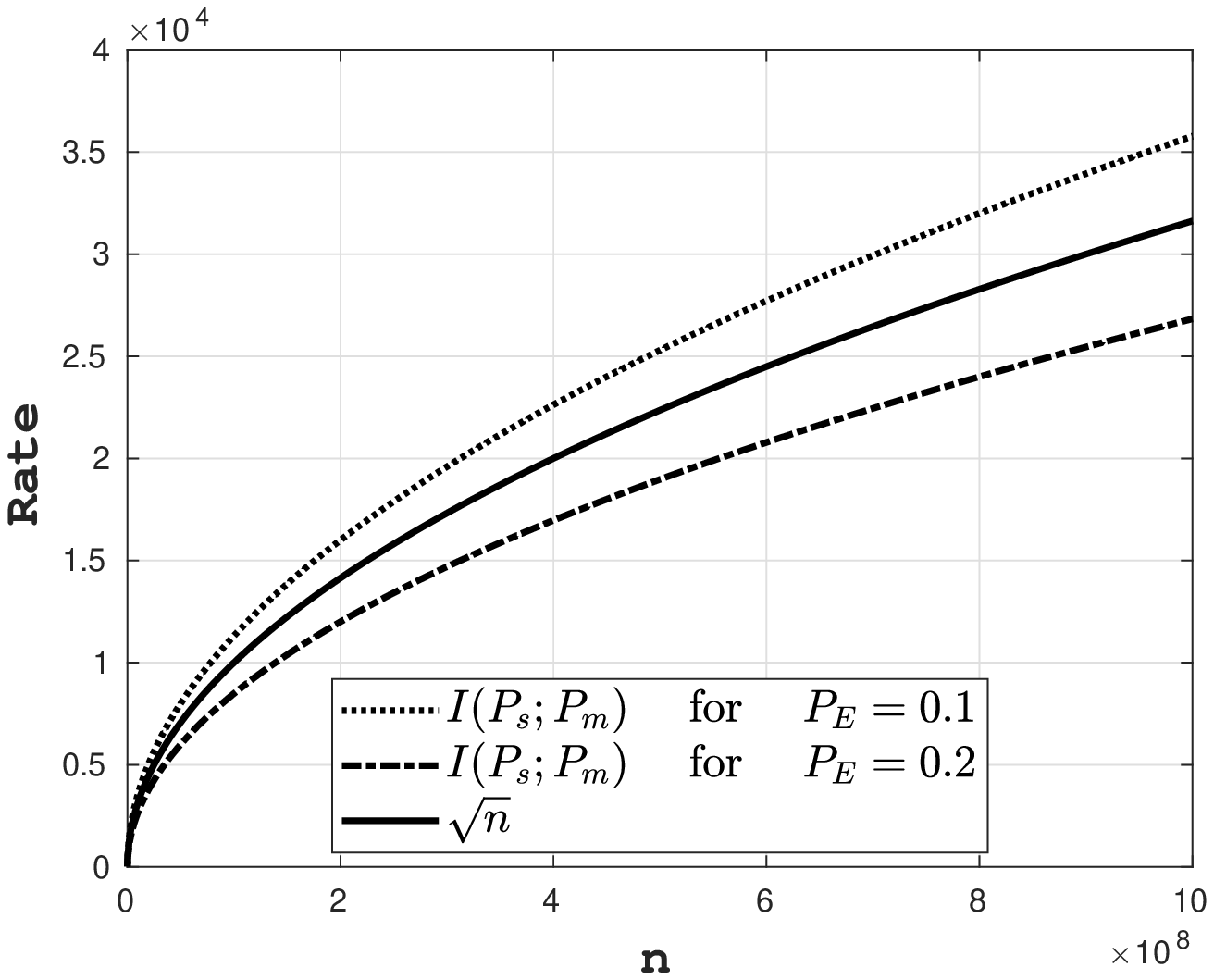}}
  \caption{Maximum achievable embedding rate $I(\StegDist;\MsgDist)$ compared to the number of cover elements ($n$) for $\AvgProbErrorEavesdropper = 0.1$ and $\AvgProbErrorEavesdropper = 0.2$. Note that, we use the lower bound for $a^*$ in \eqref{eq.result4.upper} to compute $I(\StegDist;\MsgDist)$ in~\eqref{eq.I}.}
  \label{fig:RateVsN}
\end{figure}

To illustrate the above results, and to prove the relation between $\AvgProbErrorEavesdropper$ and both \eqref{eq.main} and \eqref{eq.Isrl}, we plot in Fig.\ref{fig:RateVsN} the maximum embedding rate $I(\StegDist;\MsgDist)$ against the number of the cover elements $n$ using \eqref{eq.I} and \eqref{eq.result4.upper}. In the figure, we use the lower bound for $a^*$ in \eqref{eq.result4.upper}, i.e., we set $a^* = -W(-\frac{4\,(1-2\AvgProbErrorEavesdropper)^2}{n} -1 -i\pi)$ to compute $I(\StegDist;\MsgDist)$ in~\eqref{eq.I}. We provided the plots for $\AvgProbErrorEavesdropper = 0.1$ and $\AvgProbErrorEavesdropper = 0.2$ and we used the implementation of the wrightOmega function in \cite{wrightOmegaq} to generate the plots. From the figure, we can conclude that \eqref{eq.Wtheta}, \eqref{eq.a.approx.1}, and \eqref{eq.Isrl} are not over-approximated form of \eqref{eq.main} as we plot the actual analytic form expression in \eqref{eq.main} against $\sqrt{n}$.

\section{Practical interpretation}\label{sec:exp}
In this section, we compare the practical experimental results for the stenographic methods in \cite{quantized.gauss}, \cite{markov2}, and \cite{markov8} with our theoretically calculated limits. Specifically, we get the published results of each steganographic method that present the payload (bits per pixel) of the method against different $\AvgProbErrorEavesdropper$, where $\AvgProbErrorEavesdropper$ is obtained using different steganalysis methods. Then, we plot these published results against our theoretical limit of the payload calculated using equation \eqref{eq.main}. In our comparison, we used the BOSSbase 1.01 dataset~\cite{BOSS} where the size of each image is $512\times 512$ pixels. We set $n = 512 \times 512 = 2^{18}$ in \eqref{eq.main} to calculate the maximum achievable embedding rate that occurs in the case of independent cover elements, i.e. when each clique represents only a single pixel. We give more clarification for the relation between the achievable embedding rate and the clique size in Section~\ref{sec:discuss}.

\begin{figure}\centering
	\includegraphics[width=0.75\linewidth]{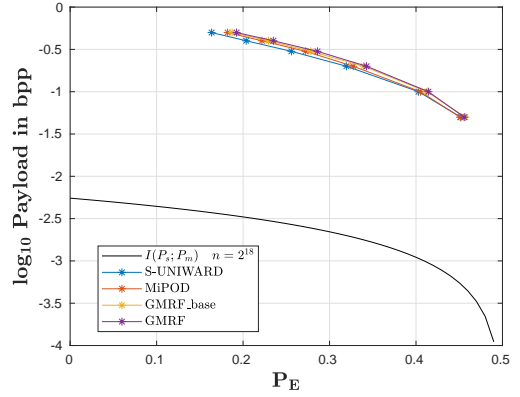}
	    \caption{Results from \cite{markov2} comparing steganographic methods: S-UNIWARD, MIPOD, GMRF\_BASE and GMRF with steganalyzer utilizing SRM feature.}\label{fig:Markov_2_T1a}
\end{figure}

\begin{figure}\centering
	\includegraphics[width=0.75\linewidth]{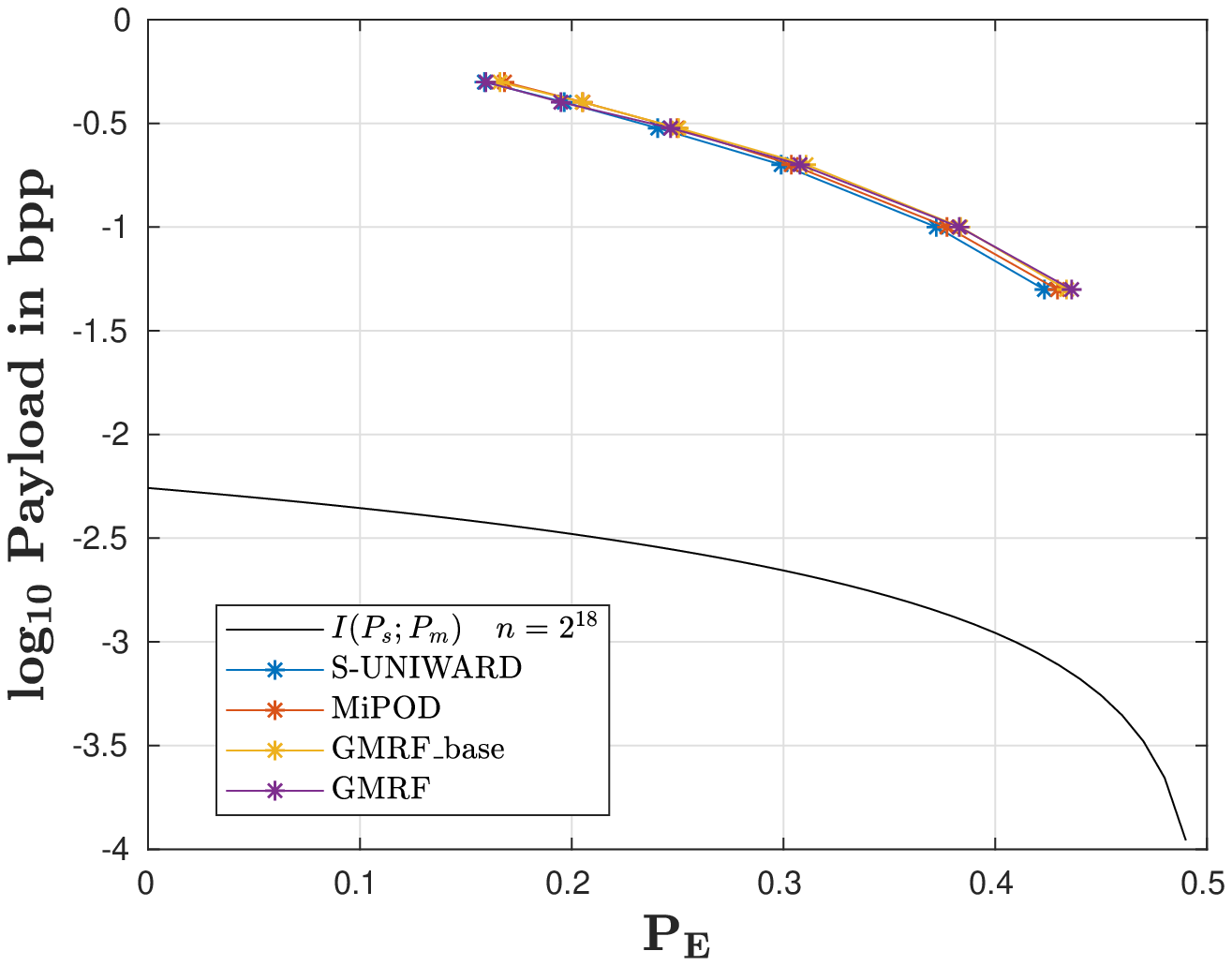}   
	    \caption{Results from \cite{markov2} comparing steganographic methods: S-UNIWARD, MIPOD, GMRF\_BASE and GMRF with steganalyzer utilizing maxSRMd2 feature.}\label{fig:Markov_2_T1b}
\end{figure}

\begin{figure}\centering
	\includegraphics[width=0.75\linewidth]{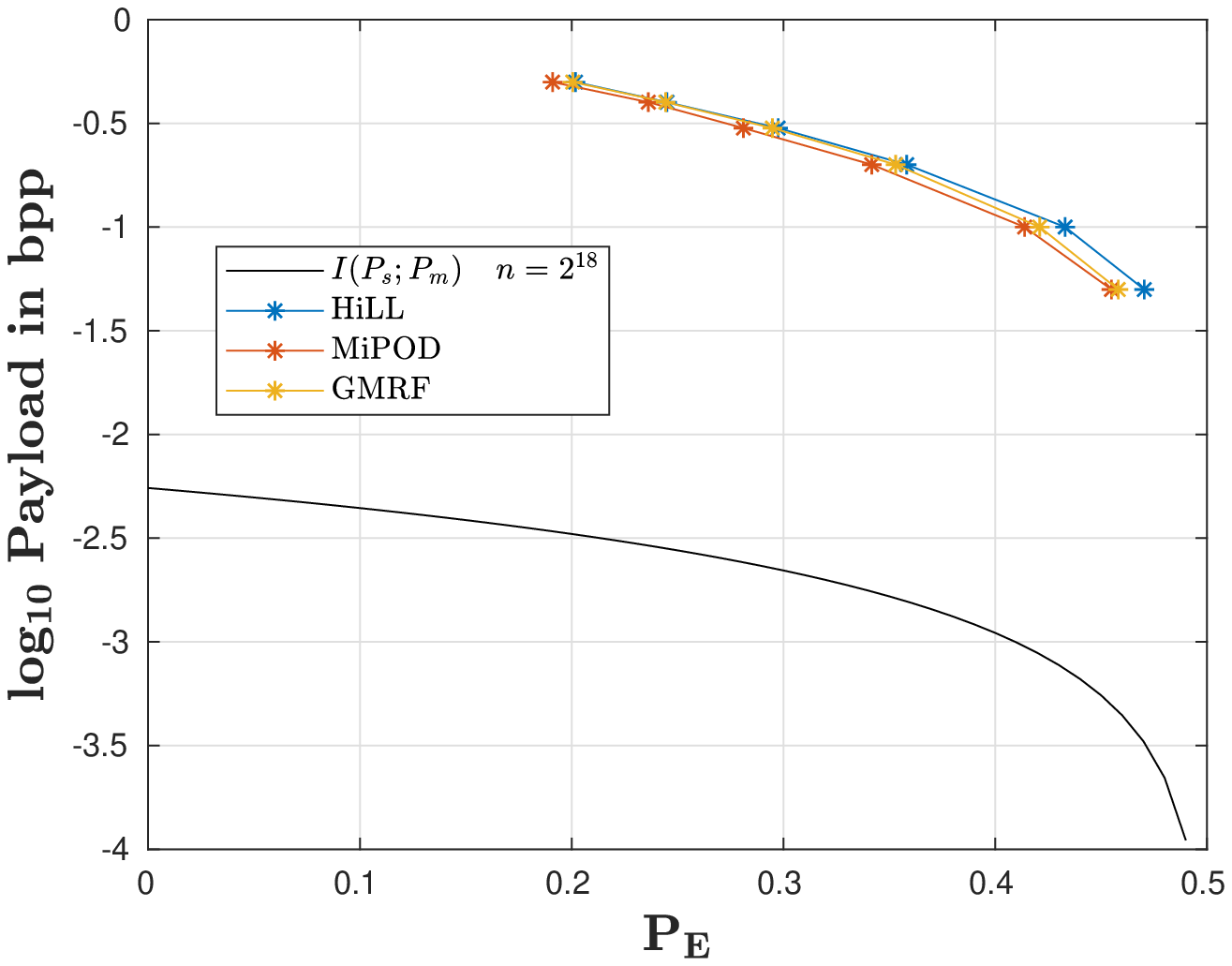}
	    \caption{Results from \cite{markov2} comparing steganographic methods: MiPOD, HILL and GMRF enhanced by low-pass-filtered-cost method with steganalyzer utilizing SRM feature.}\label{fig:Markov_2_T2a}
\end{figure}

\begin{figure}\centering
	\includegraphics[width=0.75\linewidth]{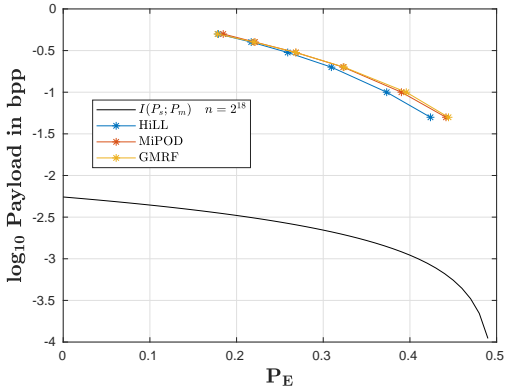}   
	    \caption{Results from \cite{markov2} comparing steganographic methods: MiPOD, HILL and GMRF enhanced by low-pass-filtered-cost method with steganalyzer utilizing maxSRMd2 feature.}\label{fig:Markov_2_T2b}
\end{figure}

\Cref{fig:Markov_2_T1a,fig:Markov_2_T1b,fig:Markov_2_T2a,fig:Markov_2_T2b} demonstrate the experimental results obtained from \cite{markov2} plotted against the theoretical upper limit for the payload in \eqref{eq.I} and \eqref{eq.result4.upper}. We set $a^*$ to its lower bound in \eqref{eq.result4.upper} to compute the theoretical upper limit for the payload against $\AvgProbErrorEavesdropper$ and we plot the payload in log-scale.
We also demonstrate in the supplementary material figures \labelcref{S-fig:Markov_8_T1a,S-fig:Markov_8_T1b,S-fig:Markov_8_T2a,S-fig:Markov_8_T2b} and figures \labelcref{S-fig:quantized_gauss_T1,S-fig:quantized_gauss_T2_HILL,S-fig:quantized_gauss_T2_MiPOD,S-fig:quantized_gauss_T2_SUNIWARD,S-fig:quantized_gauss_T3_HILL,S-fig:quantized_gauss_T3_MiPOD,S-fig:quantized_gauss_T3_SUNIWARD} for the methods in \cite{markov8} and \cite{quantized.gauss}, respectively.

It can be concluded from the figures that our theoretically calculated upper limit is relatively very small compared to the referenced practical stenographic methods. The reason for this is that there exist other steganalysis methods that can be more optimum than the methods utilized by referenced stenographic methods. In other words, the steganalysis methods used in \cite{quantized.gauss,markov2,markov8} can be regarded as non-optimal compared to the KL divergence $\mathcal{D}(\StegDist \parallel \CoverDist)$ that is used in our proof and can be regarded as the upper bound for any steganalysis detector. Consequently, using non-optimal steganographic detectors may be misleading as it can achieve a lower probability of detection error (i.e. $\AvgProbErrorEavesdropper$) when using higher embedding rates than our theoretically calculated limit. Likewise, using non-optimal embedding techniques may also be misleading as it can achieve lower embedding rates than the theoretically calculated limit with the same detection probability.

\section{Discussion}\label{sec:discuss}
As $I(\StegDist;\MsgDist)\propto\sqrt{n}$ as shown in \eqref{eq.Isrl}, the upper limit for embedding rate occurs when each clique contains a single element, i.e. when $n$ is large. This occurs in cases such as a highly textured noisy image or a motion field of a sand storm video. Meanwhile, the lowest embedding rate occurs when all the elements of the cover are fully correlated (\textit{i.e.}: the whole cover can be considered as a single clique where $n=1$), such as an image of a clear sky or a motion field of a video of the stationary scene with global camera motion. These cases imply relatively a small number of cliques. In other words, the upper limit occurs in the situation when all elements of the cover are independent, whereas the lowest one occurs in the case of the cover contains only a single clique.


Another important point is that using the information-theoretic upper bound for $I(\StegDist;\MsgDist)$ in \eqref{eq.main} is not a practical constructive approach due to the \textit{random coding argument} assumption. In other words, it does not introduce concrete guidelines for how to design steganographic methods or codebooks to achieve this theoretic upper bound. Our work introduces a proof of existence in analytical form, implying that work is still needed to construct steganographic methods to be more closed to the calculated bounds. 

\section{Conclusion}\label{sec:conc}
Through this paper, we have calculated the information-theoretic upper bound for a predefined level of security for data embedding within a generalized case for multimedia covers. Our mathematical analysis has several advantages over previous ones. First, we introduced a mathematical justification for using CMQGD as a replacement for the accurate and mathematically intractable Gibbs distribution. Second, we provided a rigorous analytic form for the SRL, not an order result. Third, the provided analytic form is calculated not only in the case when the cover is known for both sender $\Sender$ and receiver $\Receiver$, but also in the real-world case when $\Receiver$ doesn't know the exact cover, only knows the cover distribution, thanks to the achievability proof in section \ref{sec:ach}. Forth, our calculated limit is applied to all types of stegnanalytic detectors (statistical, deep-learning, feature-based, ... \textit{etc}) for any type of multimedia. Five, we have introduced the model parameters for the optimal message's codebook in an analytic form. It should be noted that for continuously distributed covers, our mathematical solutions will be exact closed-form solutions.

\section*{Appendix}
\subsection*{A. The Entropy of Multivariate Quantized Gaussian Distribution}\label{lemma.1}

\begin{lemma}\label{lem.A1}
Theorem (8.3.1) in \cite{info.th.book} defines the entropy of quantized distribution as
\begin{equation}\label{eq.ent.t.cover}
H(P) \approx h(p) + b,
\end{equation}
where $p$ is any continuous distribution, $P$ is a quantized version of $p$ with $b$ quantization bits, $H(P)$ is the entropy of the quantized distribution and $h(p)$ is the differential entropy of the continuous distribution. Thus, as the entropy of CMQGD $p_g$ is defined in \cite{limits2} as
\begin{equation}\label{eq.ent.mqgd}
h(p_g) = \frac{1}{2}\ln (2\pi e |\Sigma_g|),
\end{equation}
the entropy of both $\CoverDist$ and $\StegDist$ can be defined w.r.t their continuous versions $\CoverDist$, $\StegDist$ as
\begin{equation}\label{eq.ent.pc}
H(\CoverDist) \approx \frac{1}{2}\ln (2\pi e |\Sigma_c|) + b,
\end{equation}
\begin{equation}\label{eq.ent.ps}
H(\StegDist) \approx \frac{1}{2}\ln (2\pi e |\Sigma_s|) + b. 
\end{equation}

\end{lemma}


\subsection*{B. Additional properties about the embedding factor $a^*$}

\begin{lemma}\label{lem.A*}$a^*$ cannot be a complex number:\\
It should be noted that $a^* \in\mathbb{R}$ despite $W(.) \in \mathbb{C}$. From \cite{WrightOmega}, the relation between the Lambert W Function $\mathcal{W}(.)$ and the WrightOmega function $W(.)$ can be modelled as
\begin{equation}\label{eq.lambertw}
W(x) = \mathcal{W}_{\big \lceil \frac{\mathrm{Im}(x) - \pi}{2 \pi} \big \rceil}(e^{x}).
\end{equation}
If we set $x = -\frac{4\epsilon^2}{n} -1 -i\pi$ as in \eqref{eq.result4m}, then, $\mathrm{Im}(x) = \pi$ and therefore \eqref{eq.lambertw} can be re-written as
\begin{equation}
W(x) = \mathcal{W}_{-1}(e^{x}),
\end{equation}
which is a special case of Lambert W Function where $\mathcal{W}_{-1}(.) \in \mathbb{R}$ \cite{LambertWfunction}. Thus, $a* \in \mathbb{R}$. The same applies for $a$ in \eqref{eq.result4}.
\end{lemma}
\begin{lemma}\label{lemma.a.geq.1}$a^*$ cannot be less than 1: \\
As $\Sigma_s, \Sigma_c$ and $\Sigma_m$ are positive semi-definite matrices, then from \eqref{eq.sigmam2} $a$ and $a^*$ must be greater than 1, with equality when no embedding occurs, as explained in \eqref{eq.result2.a} and \eqref{eq.sigmam2}. Thus:
\begin{equation}\label{eq.agt1}
a \geq 1.
\end{equation}
\end{lemma}

\begin{lemma}\label{lamma.a.pe}Relation between $\AvgProbErrorEavesdropper$ and $a^*$:\\
Using \eqref{eq.pe_pd} through \eqref{eq.klv} and \eqref{eq.c1}, we have:
\begin{equation}\label{eq.dpe}
2\,(1-2\AvgProbErrorEavesdropper)^2 \leq \mathcal{D}(\StegDist \parallel \CoverDist) \leq 2\epsilon^2,
\end{equation}
Then, as the WrightOmega function in the form $-W(-\gamma-1-i\pi)$ is monotonically increasing with $\gamma$ \cite{WrightOmega}, where $\gamma$ is a positive constant, we can prove that
\begin{align}\label{eq.result4.upper}
-W(-\frac{4\,(1-2\AvgProbErrorEavesdropper)^2}{n} -1 -i\pi) &\leq a \leq -W(-\frac{4\epsilon^2}{n} -1 -i\pi) \nonumber \\
-W(-\frac{4\,(1-2\AvgProbErrorEavesdropper)^2}{n} -1 -i\pi) &\leq a \leq a^*,
\end{align}
by utilizing equations \eqref{eq.result4}, \eqref{eq.result4m} and \eqref{eq.dpe}.
\end{lemma}
\bibliographystyle{IEEEtran}
\bibliography{IEEEabrv,References}



\end{document}


\maketitle
\begin{figure}[ht]\centering
	\includegraphics[width=0.95\linewidth]{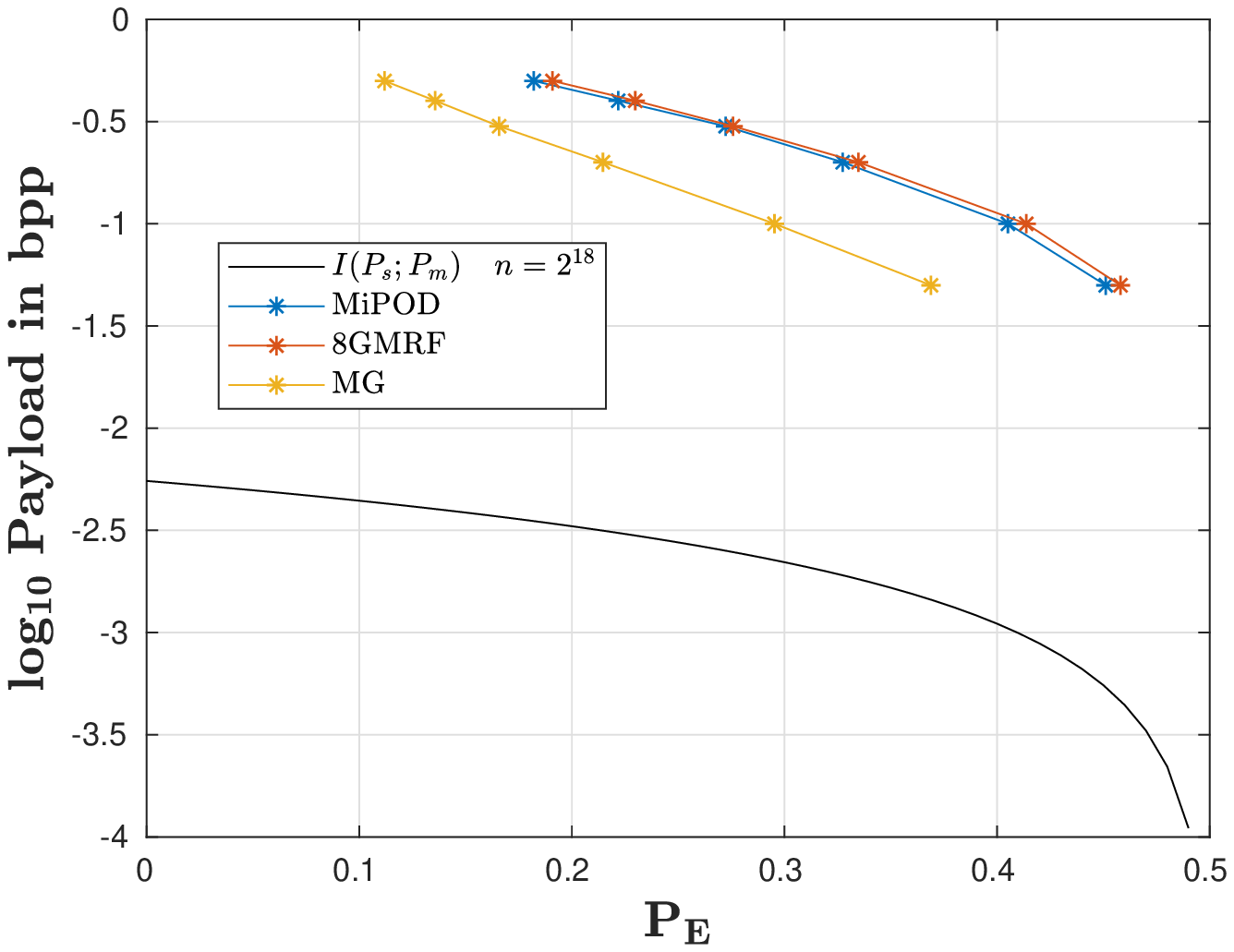}
	    \caption{Results from \cite{markov8} comparing steganographic methods: MiPOD, 8-GMRF and MG with steganalyzer utilizing SRM feature.}\label{fig:Markov_8_T1a}
\end{figure}
\begin{figure}\centering
	\includegraphics[width=0.95\linewidth]{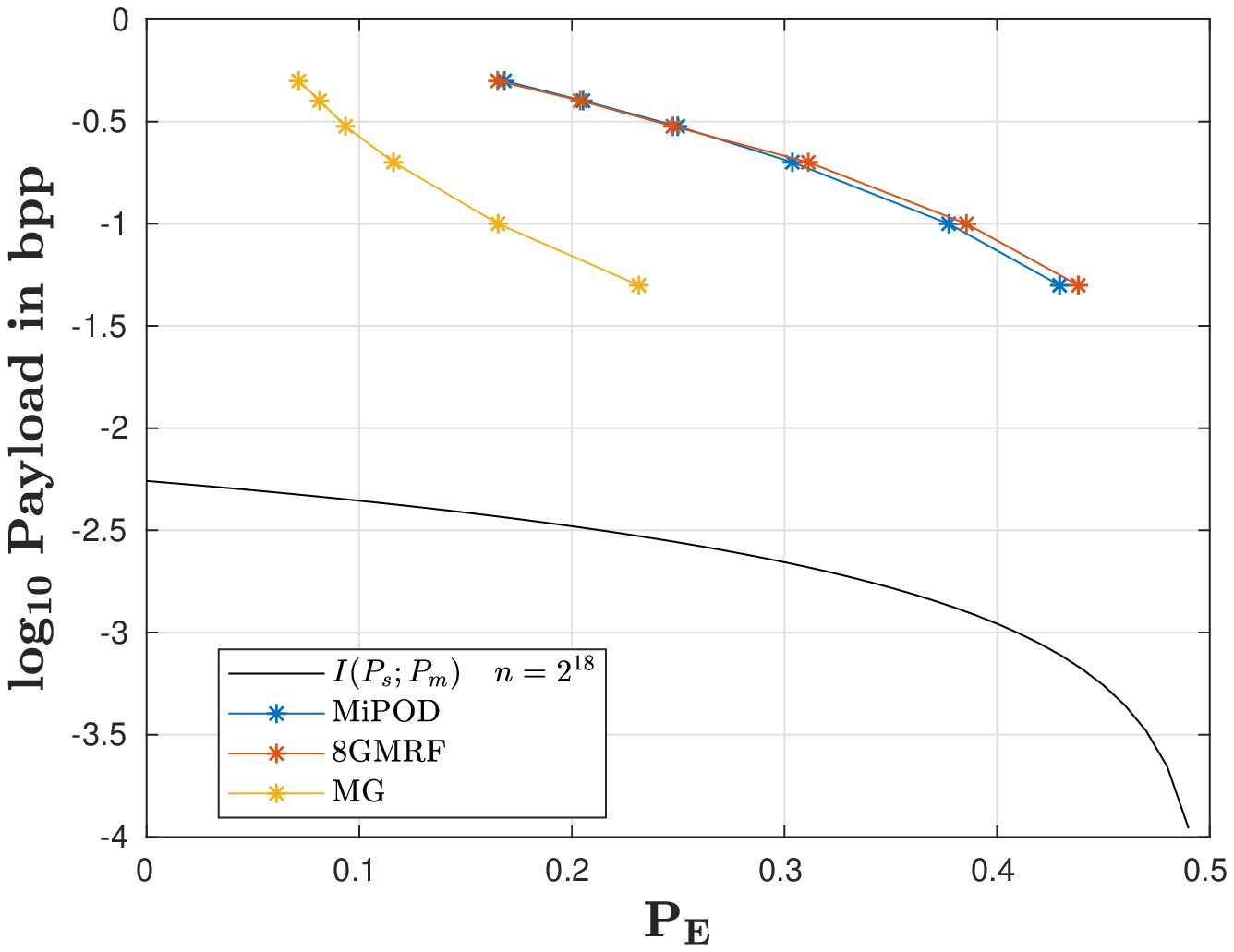}   
	    \caption{Results from \cite{markov8} comparing steganographic methods: MiPOD, 8-GMRF and MG with steganalyzer utilizing maxSRMd2 feature.}\label{fig:Markov_8_T1b}
\end{figure}
\begin{figure}\centering
	\includegraphics[width=0.95\linewidth]{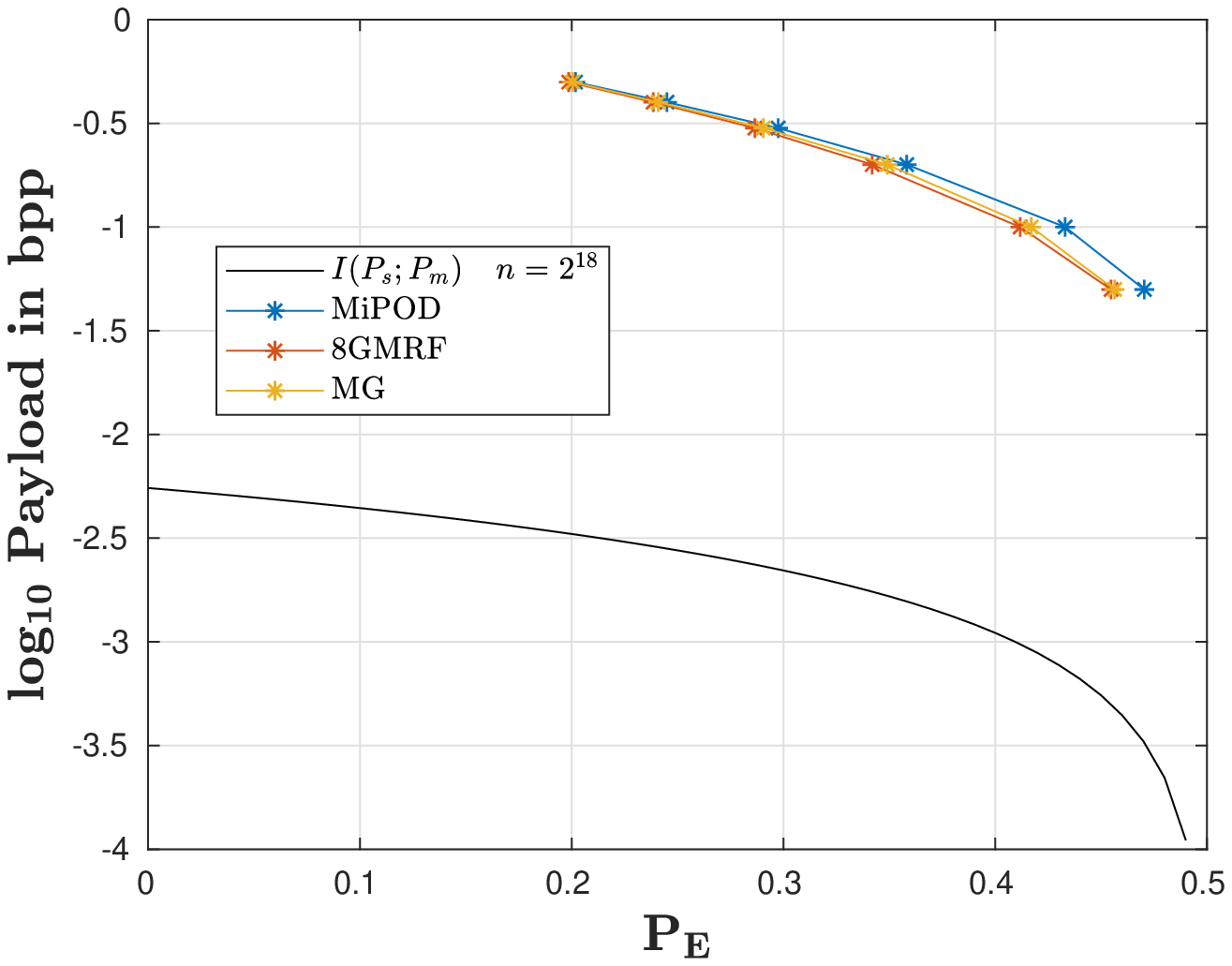}
	    \caption{Results from \cite{markov8} comparing steganographic methods: HiLL, MiPOD and 8-GMRF enhanced by filtered cost using ensemble 1.0 classifier with steganalyzer utilizing SRM feature.}\label{fig:Markov_8_T2a}
\end{figure}
\begin{figure}\centering
	\includegraphics[width=0.95\linewidth]{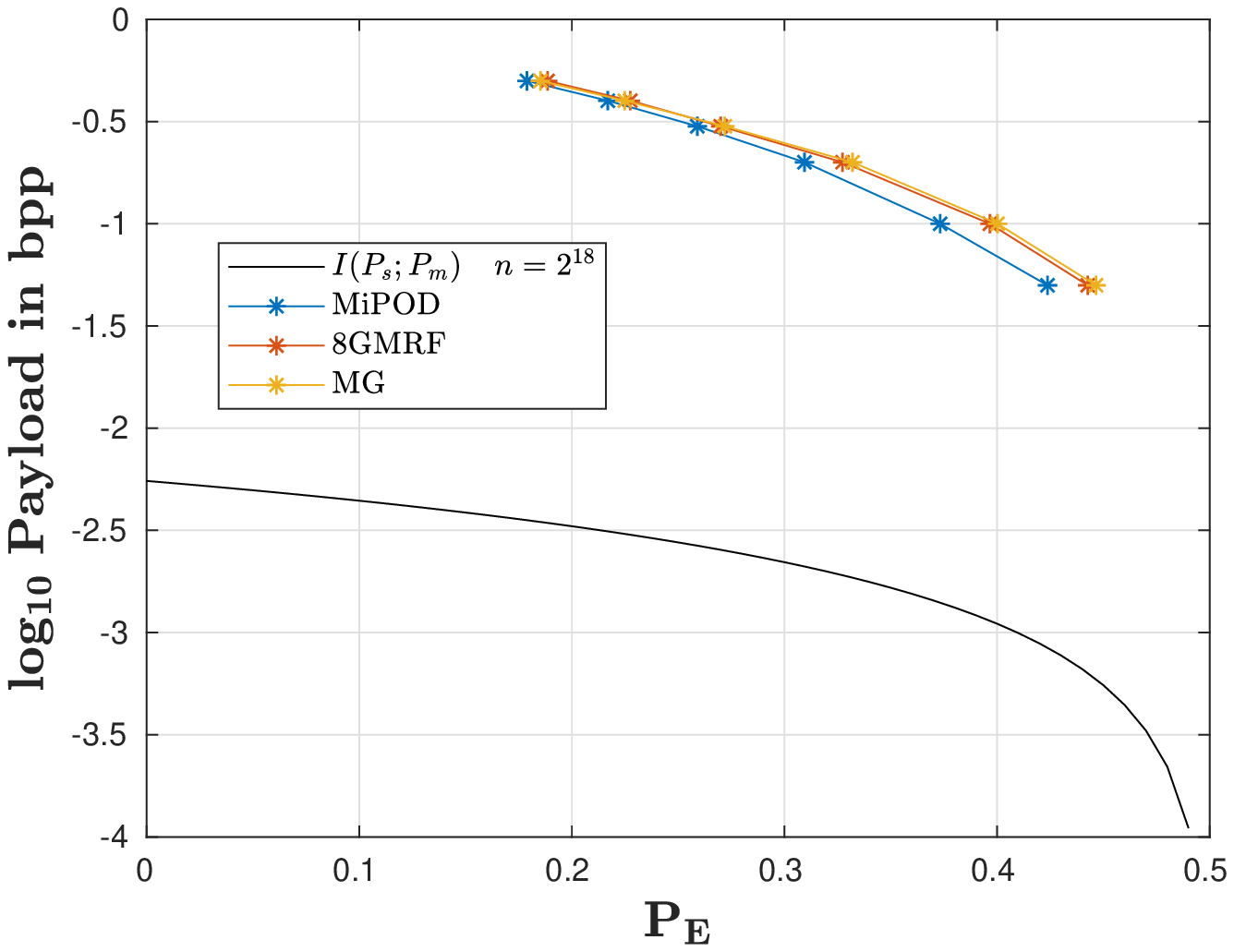}   
	    \caption{Results from \cite{markov8} comparing steganographic methods: HiLL, MiPOD and 8-GMRF enhanced by filtered cost using ensemble 1.0 classifier with steganalyzer utilizing maxSRMd2 feature.}\label{fig:Markov_8_T2b}
\end{figure}
\begin{figure}
  \centerline{\includegraphics[width=0.95\linewidth]{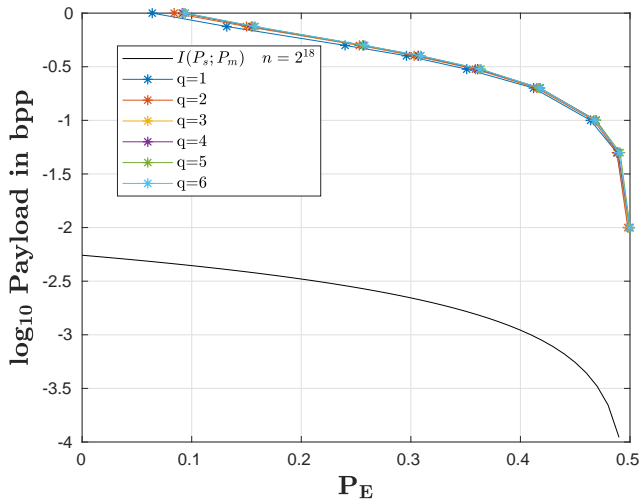}}
  \caption{Comparing our theoretically calculated limit with results of the Gaussian version of the HILL algorithm in \cite{quantized.gauss} with different $q$ values in a $(2q+1)$-ary embedding with steganalyzer utilizing maxSRMd2 feature.}
  \label{fig:quantized_gauss_T1}
\end{figure}
\begin{figure}\centering
\includegraphics[width=0.95\linewidth]{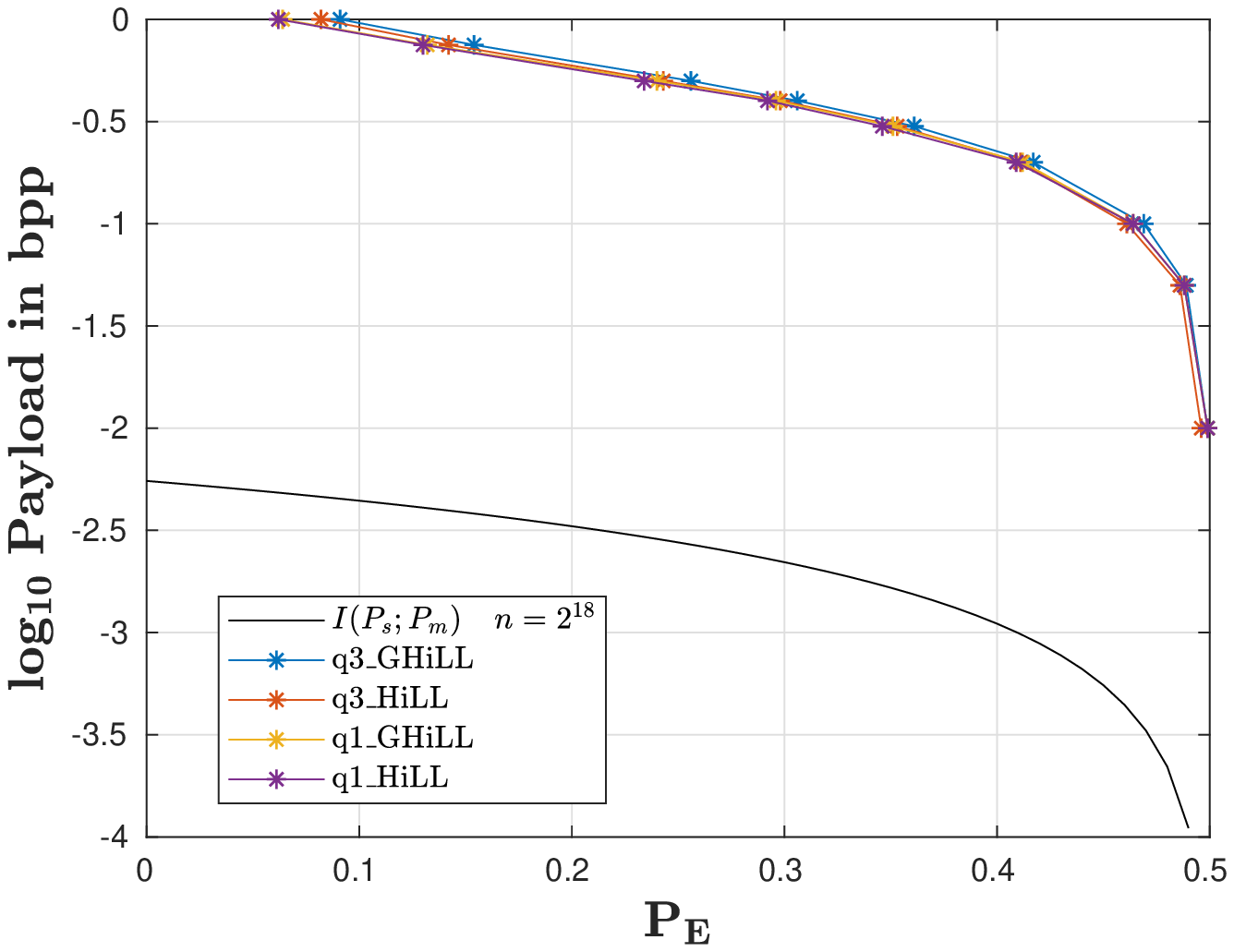}
    \caption{Results from \cite{quantized.gauss} comparing HILL steganographic method and its modified Gaussian version with $q=1,3$ with steganalyzer utilizing maxSRMd2 feature.}
    \label{fig:quantized_gauss_T2_HILL}
\end{figure}
\begin{figure}\centering
\includegraphics[width=0.95\linewidth]{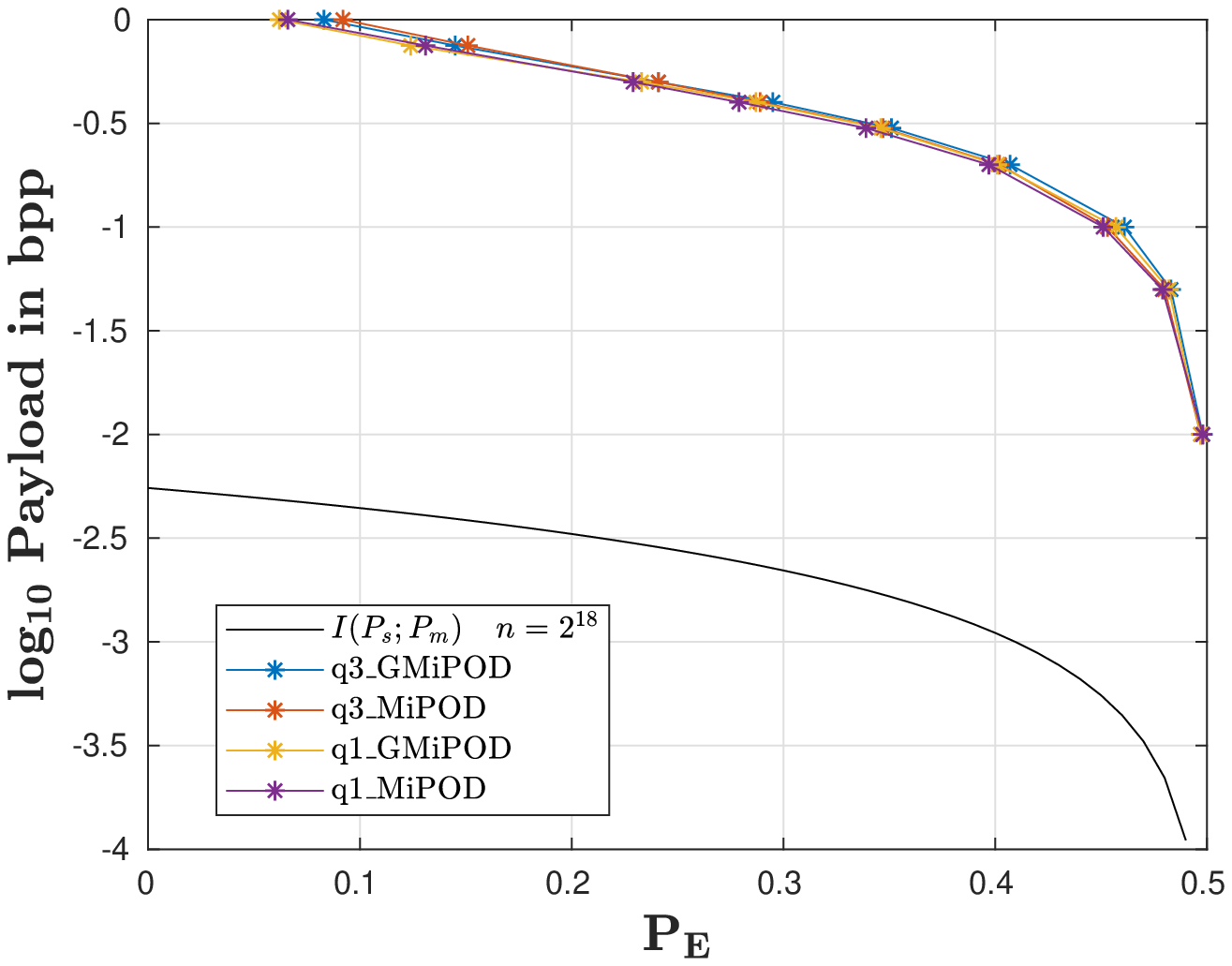}   
    \caption{Results from \cite{quantized.gauss} comparing MiPOD steganographic method and its modified Gaussian version with $q=1,3$ with steganalyzer utilizing maxSRMd2 feature.}
    \label{fig:quantized_gauss_T2_MiPOD}
    \end{figure}
\begin{figure}\centering
\includegraphics[width=0.95\linewidth]{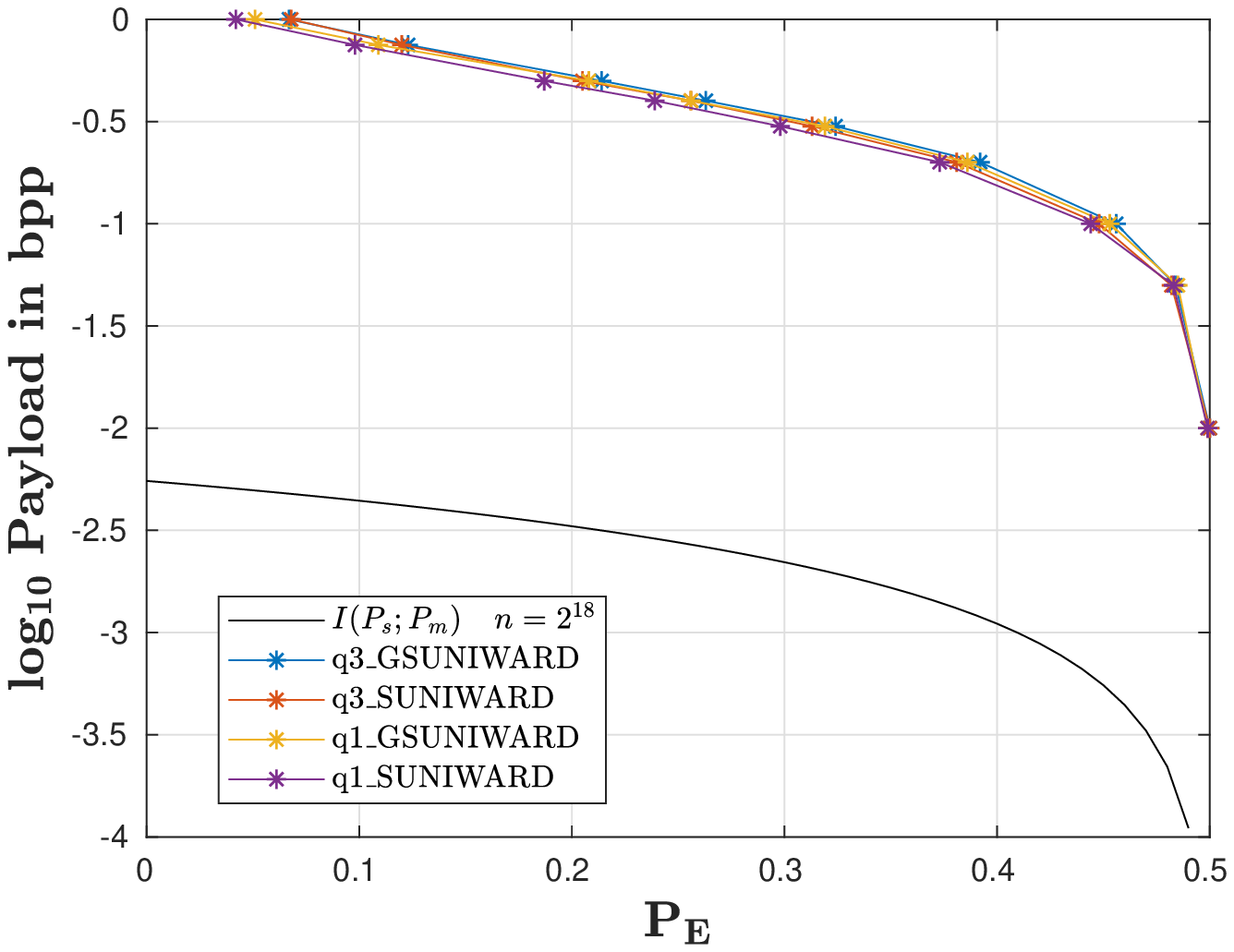}
    \caption{Results from \cite{quantized.gauss} comparing SUNIWARD steganographic method and its modified Gaussian version with $q=1,3$ with steganalyzer utilizing maxSRMd2 feature.}
    \label{fig:quantized_gauss_T2_SUNIWARD}
\end{figure}
\begin{figure}\centering
\includegraphics[width=0.95\linewidth]{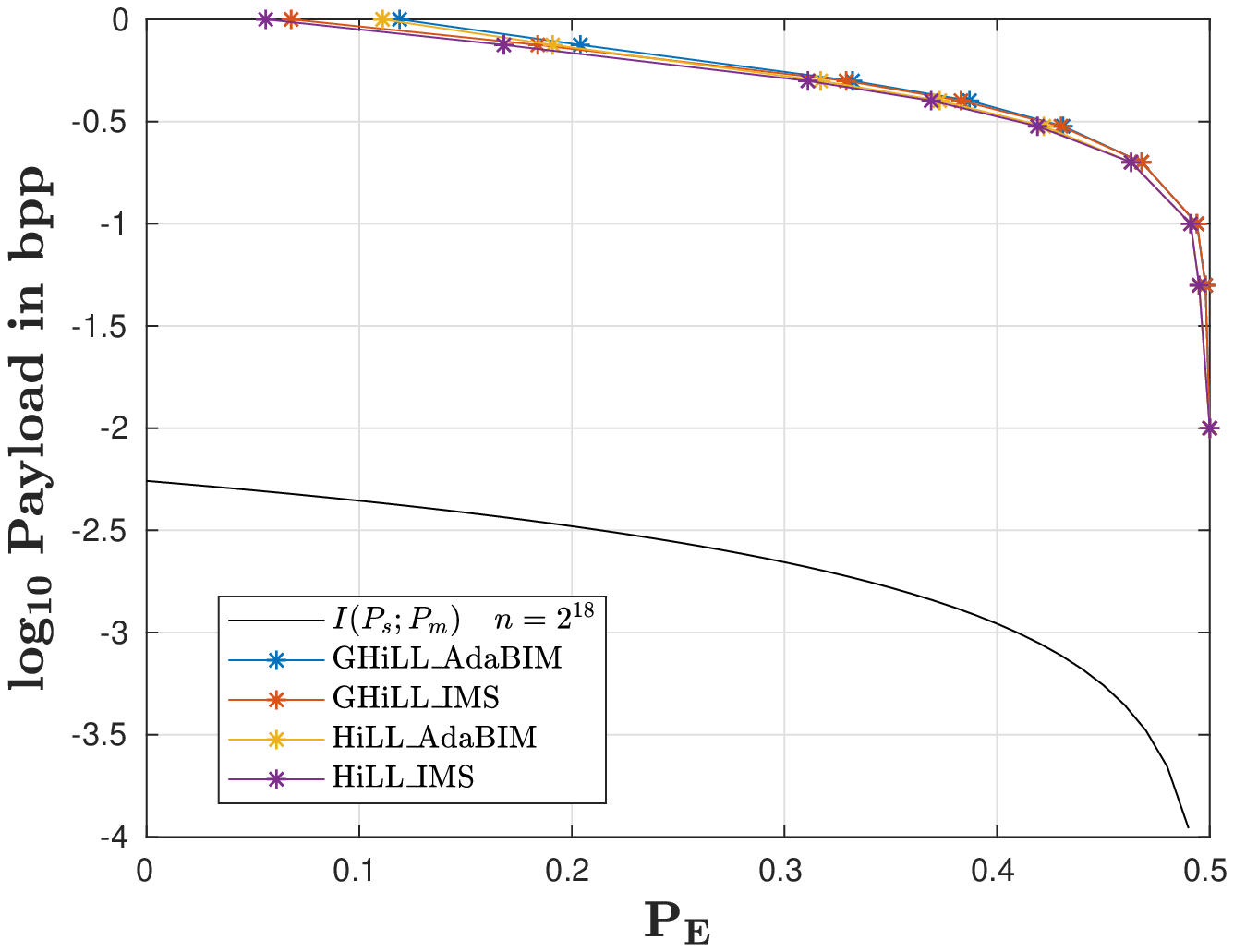}
    \caption{Results from \cite{quantized.gauss} comparing HILL steganographic method and its modified Gaussian version with two different batching strategies, $IMS$ with batch size 128 and $AdaBIM$ with adaptive batch size against steganalyzer utilizing maxSRMd2 feature.}
    \label{fig:quantized_gauss_T3_HILL}
\end{figure}
\begin{figure}\centering
\includegraphics[width=0.95\linewidth]{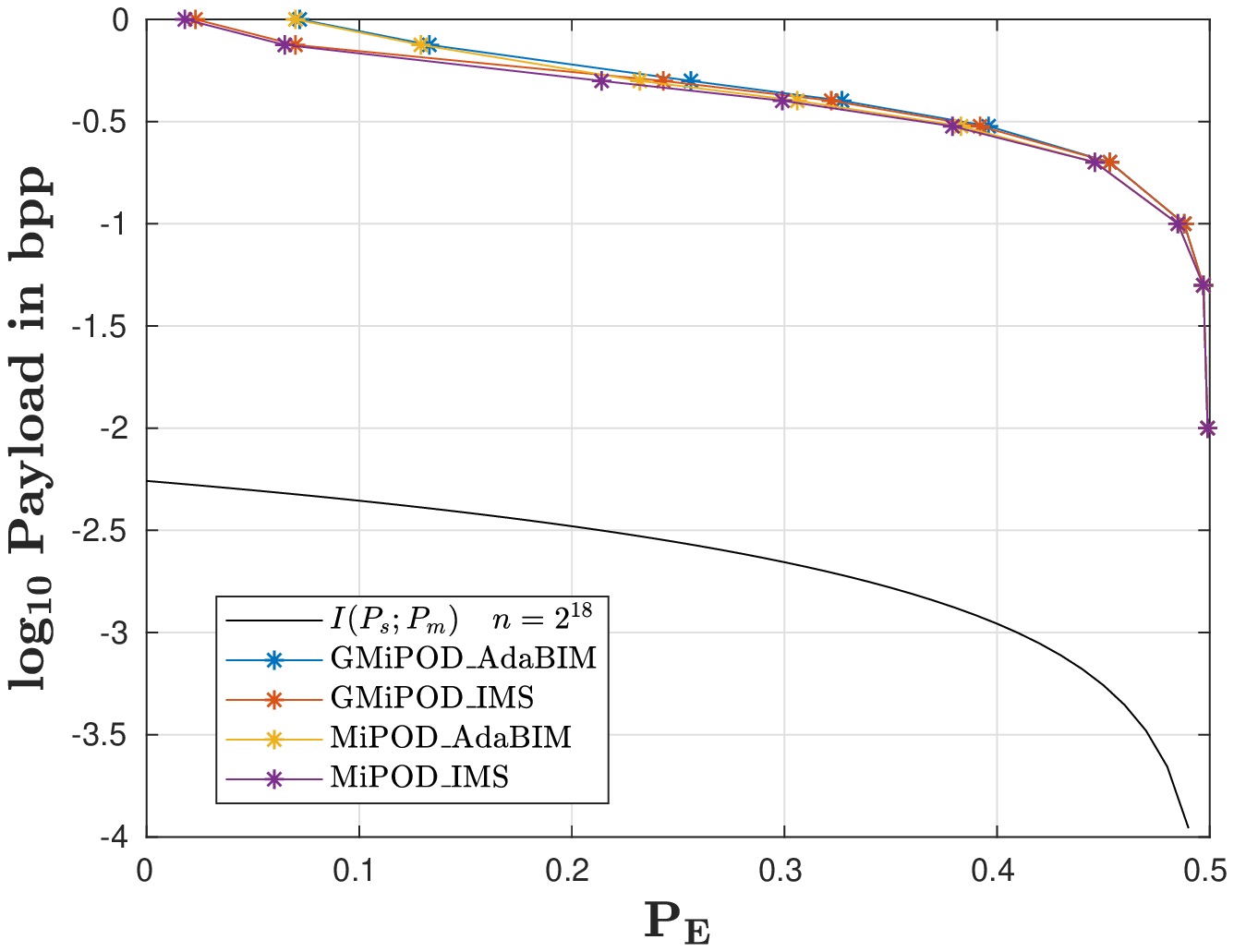}   
    \caption{Results from \cite{quantized.gauss} comparing HILL steganographic method and its modified Gaussian version with two different batching strategies, $IMS$ with batch size 128 and $AdaBIM$ with adaptive batch size against steganalyzer utilizing maxSRMd2 feature.}
    \label{fig:quantized_gauss_T3_MiPOD}
    \end{figure}
\begin{figure}\centering
\includegraphics[width=0.95\linewidth]{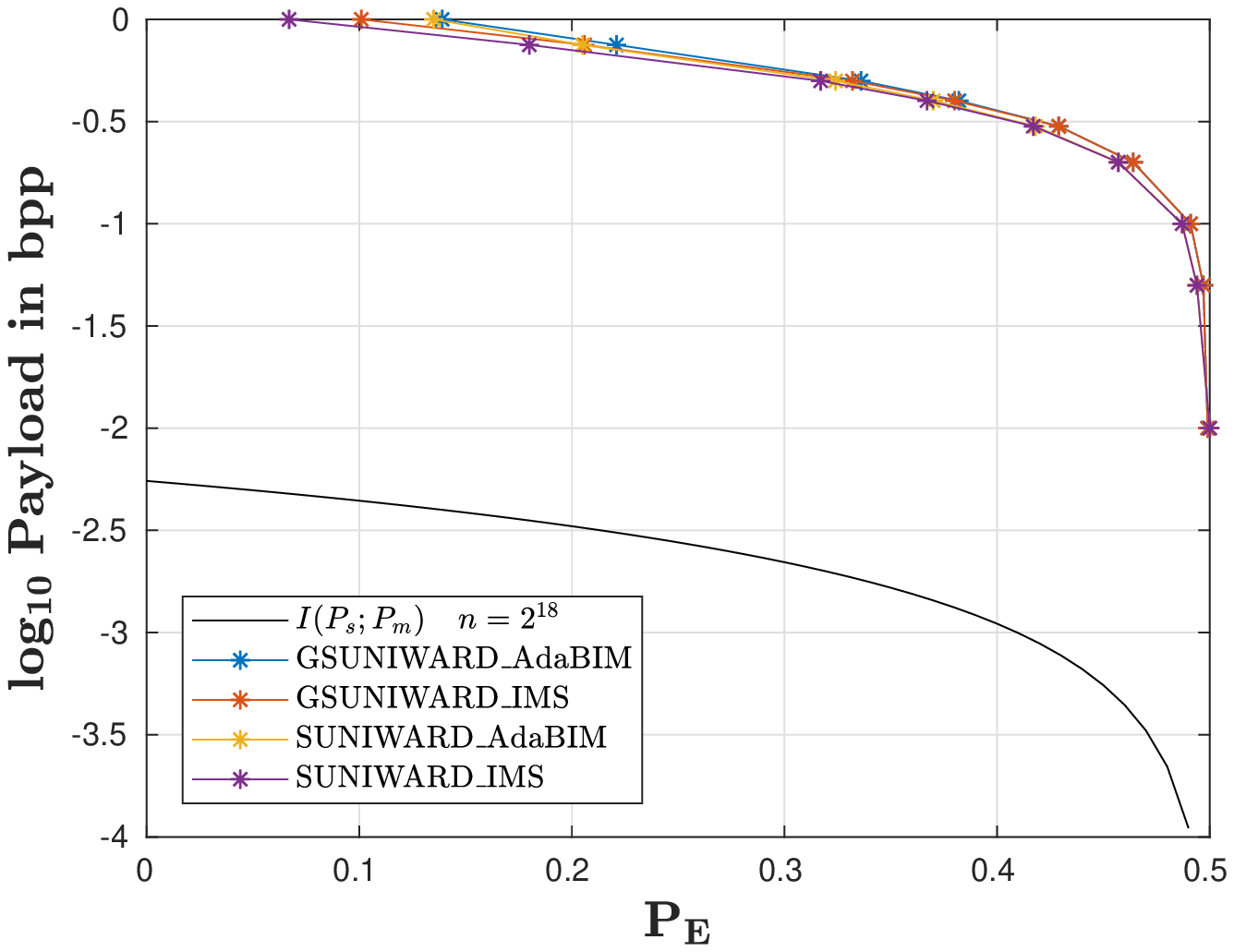}
    \caption{Results from \cite{quantized.gauss} comparing HILL steganographic method and its modified Gaussian version with two different batching strategies, $IMS$ with batch size 128 and $AdaBIM$ with adaptive batch size against steganalyzer utilizing maxSRMd2 feature.}
    \label{fig:quantized_gauss_T3_SUNIWARD}
\end{figure}

In this supplementary document we demonstrate an additional experimental results from \cite{markov8} and \cite{quantized.gauss} plotted against our theoretical upper bound. Figures \labelcref{fig:Markov_8_T1a,fig:Markov_8_T1b,fig:Markov_8_T2a,fig:Markov_8_T2b} demonstrates the results obtained from \cite{markov8} and figures \labelcref{fig:quantized_gauss_T1,fig:quantized_gauss_T2_HILL,fig:quantized_gauss_T2_MiPOD,fig:quantized_gauss_T2_SUNIWARD,fig:quantized_gauss_T3_HILL,fig:quantized_gauss_T3_MiPOD,fig:quantized_gauss_T3_SUNIWARD}   are obtained from \cite{quantized.gauss}.


\clearpage
\bibliographystyle{IEEEtran}
\bibliography{IEEEabrv,References}